\documentclass[preprint,authoryear,12pt]{elsarticle}

\usepackage{lscape, epsfig}

\usepackage{amssymb}

\usepackage[ps2pdf,%
a4paper=true,%
breaklinks=true,%
colorlinks=true,%
pdfauthor={First Author et al.},%
pdftitle={Template for manuscripts in Advances in Space Research}%
]{hyperref}

\journal{Advances in Space Research}

\begin{document}

\begin{frontmatter}



\title{CCD $UBV$ Photometry and Kinematics of the Open Cluster NGC~225}

\author[Ist]{Sel\c cuk Bilir\corref{cor}}\ead{sbilir@istanbul.edu.tr}
\author[Ist]{Z. Funda Bostanc\i}
\author[Ens]{Talar Yontan}
\author[Ist]{Tolga G\"uver}
\author[Akd]{Volkan Bak\i \c s}
\author[Ist]{Tansel Ak}
\author[Ist]{Serap Ak}
\author[Mas]{Ernst Paunzen}
\author[Akd]{Zeki Eker}

\address[Ist]{Istanbul University, Faculty of Science, Department 
of Astronomy and Space Sciences, 34119 University, Istanbul, Turkey}
\address[Ens]{Istanbul University, Graduate School of Science and Engineering, 
Department of Astronomy and Space Sciences, 34116, Beyaz\i t, Istanbul, Turkey}
\address[Akd]{Department of Space Sciences and Technologies, Faculty of 
Sciences, Akdeniz University, Antalya 07058, Turkey}
\address[Mas]{Department of Theoretical Physics and Astrophysics, 
Masaryk University, Kotl\'a\u rsk\'a 2, 611 37 Brno, Czech Republic}
\cortext[cor]{Corresponding author}

\begin{abstract}
We present the results of CCD $UBV$ photometric and spectroscopic 
observations of the open cluster NGC~225. In order to determine the 
structural parameters of NGC~225, we calculated the stellar density 
profile in the  cluster's field. We estimated the probabilities of 
the stars being physical members of the cluster using the existing 
astrometric data. The most likely members of the cluster were used 
in the determination of the astrophysical parameters of the cluster. 
We calculated the mean radial velocity of the cluster as 
$V_{r}=-8.3\pm 5.0$ km s$^{-1}$ from the optical spectra of eight 
stars in the cluster's field. Using the $U-B$ vs $B-V$ two-colour 
diagram and UV excesses of the F-G type main-sequence stars, the 
reddening and metallicity of NGC~225 were inferred as 
$E(B-V)=0.151\pm 0.047$ mag and $[Fe/H]=-0.11\pm 0.01$ dex, 
respectively. We fitted the colour-magnitude diagrams of NGC~225 
with the PARSEC isochrones and derived the distance modulus, distance 
and age of the cluster as $\mu_{V}=9.3\pm 0.07$ mag, $d=585\pm 20$ pc 
and $t=900\pm 100$ Myr, respectively. We also estimated the galactic 
orbital parameters and space velocity components of the cluster and 
found that the cluster has a slightly eccentric orbit of 
$e=0.07\pm 0.01$ and an orbital period of $P_{orb}= 255\pm 5$ Myr. 
\end{abstract}


\end{frontmatter}
\parindent=0.5 cm


\section{Introduction}
NGC~225 ($\alpha_{2000.0}=00^{h}43^{m}39^{s}, 
\delta_{2000.0}=+61^{\circ}46^{'}30^{''}$; $l=122^{\circ}.01$, 
$b=-1^{\circ}.08$; WEBDA database\footnote{webda.physics.muni.cz}) is 
a sparsely populated and not a well-studied open  cluster. Its age 
determined by \cite{LMM91} from photographic measurements and a 
recently revised study by \cite{SMK06} do not agree with each 
other. The limited number of observations, which are from older 
photographic measurements and some 2MASS data, motivated us to observe 
and study NGC 225 by contemporary CCD technology at optical wavelengths.   

Proper motions and approximate photographic visual magnitudes of the stars 
in the field of NGC~225  were first given by \cite{Lee26}. First precise 
$UBV$ photographic  and photoelectric  measurements  of  the
cluster were performed by  \cite{Hoagetal61}, who also constructed $V$
vs $B-V$ colour-magnitude (CMD) and $U-B$ vs $B-V$ two-colour diagrams
(TCD).   \cite{Johetal61} measured  the reddening  for the  cluster as
$E(B-V)=   0.29$    mag   using   the    data   presented   by
\cite{Hoagetal61}. \cite{Svo62} determined the  spectral classes for a
number  of  stars  in  the  field   of  the  cluster,  for  which  the
photoelectric magnitudes were already  given by \cite{Hoagetal61}, and
measured the reddening,  the distance modulus and the distance  of the 
cluster as $E(B-V)=0.29$ mag, $(m-M)=9.0$ mag and $d=630$  pc, respectively.
\cite{HoApp65} measured $H_{\gamma}$  equivalent widths  of the
brighter  stars in  the  field of  the  cluster photoelectrically  and
determined their  spectral classes.  The distance modulus of the cluster 
were estimated as $(m-M)= 9.1$  mag. But, later when \cite{BeFen71} catalogued 
open  clusters its distance was re-established again, where the reddening, 
uncorrected the distance modulus, the distance and the apparent diameter of the 
cluster were given as  $E(B-V)=0.29$~mag,   $(m-M)=9.87$~mag, $d=630$~pc and
$D=14$~arcmin, respectively.  \cite{LMM91} investigated NGC~225 in detail
and determined  28 probable member  stars of the cluster  according to
the proper motions measured in their study. They had used the photographic plates 
taken in $B$ and $V$ bands for estimating its age, reddening  and distance as
$t=  120$~Myr,  $E(B-V)=0.25\pm 0.08$~mag and $d=525\pm 73$~pc,
respectively. Almost one and a half decades later, the cluster was studied by
\cite{SMK06}, who re-estimated the cluster parameters using
photographic $UBV$ and Two Micron All Sky Survey's $JHK_s$ photometry
\citep[$2MASS$,][]{Sktr2006}. \cite{SMK06} estimated the age of the
cluster differently as  $t=0.5-10$~Myr and argued that its age is not 120~Myr as
already suggested  by \cite{LMM91}. Strengthening this conclusion 
they proposed that two Herbig Be stars with $H_{\alpha}$ 
emission, dust lanes and nebulosity exist in the vicinity of the cluster implying 
possible results of recent star formation.

In this  study, we report our conclusions extracted from our CCD $UBV$ 
observations of NGC~225. Our report includes the mean radial velocity (RV) 
of the cluster measured from low-resolution optical spectra of some brighter 
stars of NGC~225.  From  these data,  we have calculated the  cluster's 
astrophysical and kinematical  parameters.  Using the  proper motions of 
the stars, we have estimated their probabilities  of being physical members 
of the cluster. We find the reddening and metallicity of the cluster
following two independent  methods. Its distance modulus  and age were
inferred by  fitting stellar  isochrones to  the observed  CMDs, while
keeping      the      reddening     and      metallicity      constant
\citep{Bilir10,Yonetal2015,Bosetal2015,Aketal2016}.

We have summarized the observations and the data reductions in Section 2. 
The CMDs, structural  parameters  of  NGC~225,  and  the  membership
probabilities of the stars in the cluster field were presented in 
Section 3. In Section 4, we measure  the astrophysical  parameters of 
the cluster. Section 5 discusses the results.

\section{Observations}
\subsection{Photometry}

CCD $UBV$ observations  of NGC~225 were carried out on  18th July 2012
using  the  1m  Ritchey-Chr\'etien  telescope (T100)  located  at  the
T\"UB\.ITAK National Observatory (TUG)\footnote{www.tug.tubitak.gov.tr} 
in Bak{\i}rl{\i}tepe, Antalya/Turkey. A composite $V$-band image taken 
with a total exposure time of 30~s is shown in Fig.~1. Fig. 1 also 
shows an image obtained from 2MASS with the same field of view. Dust 
lanes surrounding BD~+61~154 to the upper side of the 2MASS image can 
be seen while it is not present in our $V$-band image. Since CCD $UBV$ 
images of NGC~225,  NGC~6811 and NGC~6866 were  taken by T100 in the 
same night, the details of the observations and photometric reductions 
can be found in \cite{Yonetal2015}  and  \cite{Bosetal2015}, where  
the  photometric analyses of NGC~6811 and  NGC~6866 are discussed, 
respectively. Below, a brief summary is given.

\begin{figure}
\begin{center}
\includegraphics[trim=4cm 2cm 4cm 2cm, clip=true, scale=0.80]{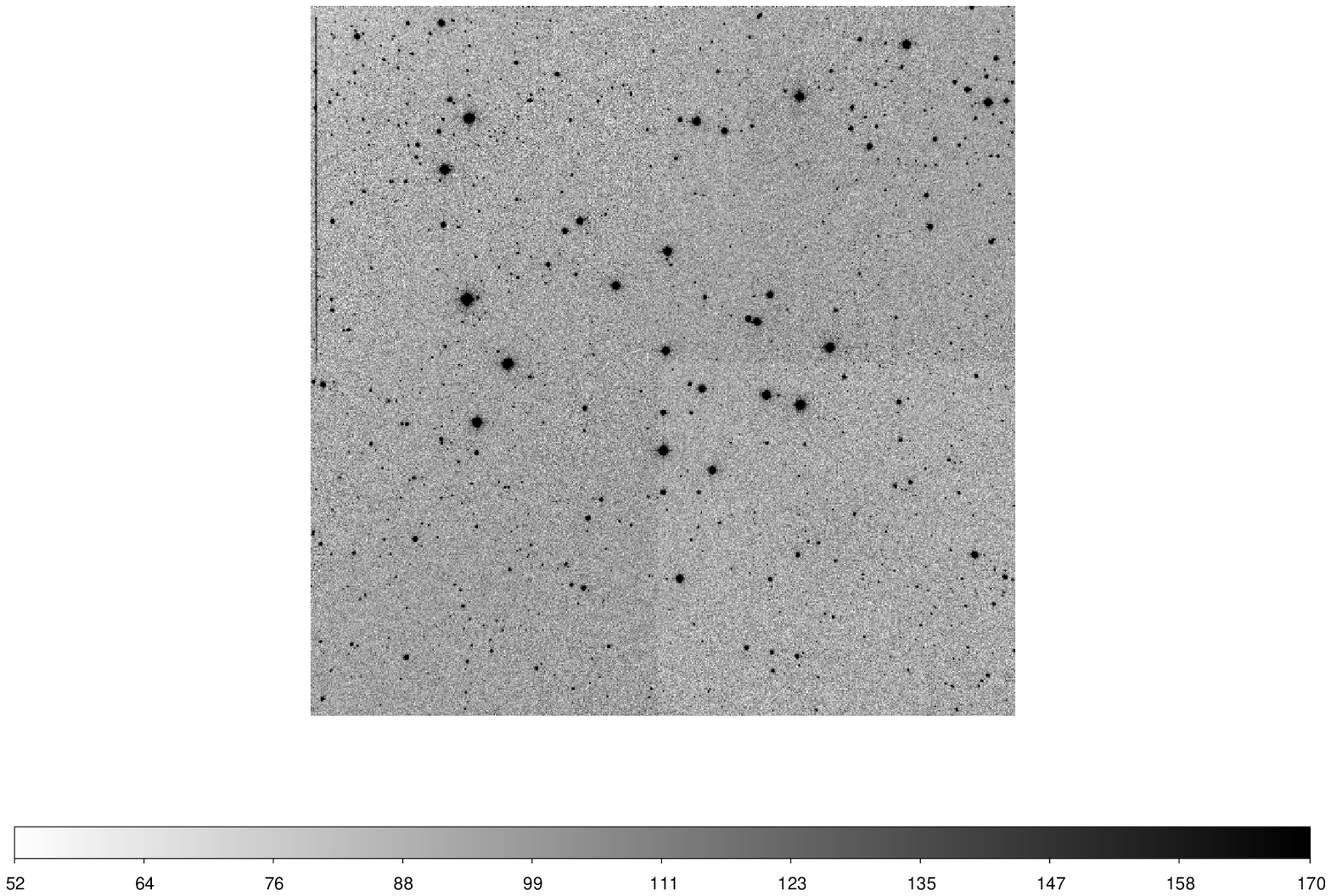}
\includegraphics[trim=0cm 4cm 0cm 0cm, clip=true, scale=0.40]{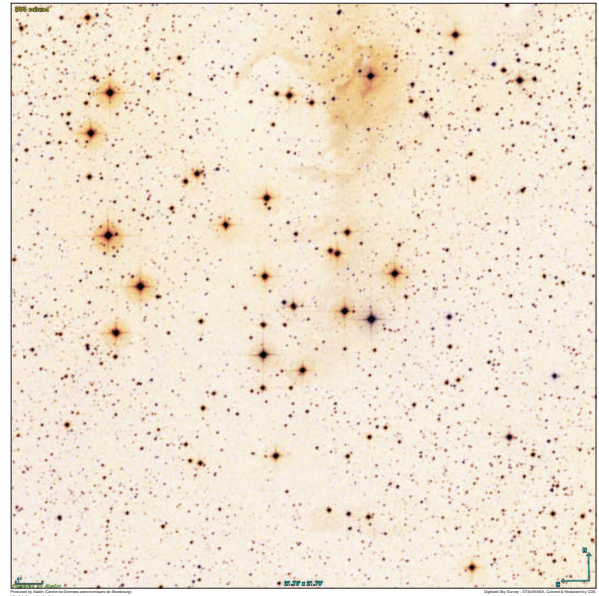}
\caption[]  {An  inverse coloured  composite $V$-band image of
  NGC~225 ({\em upper}) obtained with  T100 telescope  of the  T\"UB\.ITAK  
  National Observatory. The integral exposure time and the field of view are 30
  sec and about 21$\times$21 arcmin,  respectively (North top and East
  left). A 2MASS composite image with the same size is also presented ({\em lower}).}
\end{center}
\end{figure}

\begin{table}
\caption{Exposure times for each passband. $N$ denotes the number of exposure.}
\begin{center}
\begin{tabular}{cc}
\hline\hline
Filter & Exp. time (s)$\times$N  \\
 \hline
$U$ & 180$\times$3 , 30$\times$3 \\
$B$ & ~30$\times$3 , ~3$\times$3 \\
$V$ & ~10$\times$3 , ~1$\times$3 \\
\hline\hline
\end{tabular}
\end{center}
\end{table}

Images  of the  cluster's field  were  acquired with the  $UBV$
filters with  short and  long exposures to  cover the  widest possible
flux range. The night was moderately photometric with a mean seeing of
1$''$.5. Exposure  times for  each passband are  given in  Table~1. 
IRAF\footnote{IRAF is  distributed   by  the  National  Optical
Astronomy Observatories.}, PyRAF\footnote{PyRAF is a product of
the Space Telescope  Science Institute, which is  operated by AURA
for   NASA},  and   astrometry.net\footnote{http://astrometry.net}  
\citep{Lang09} were used for pre-reduction processes of images and
transforming  the pixel  coordinates  of the  objects identified  in
frames to equatorial coordinates.  The aperture photometry packages
of IRAF  were utilized to  measure the instrumental magnitudes  of the
standard  stars.    From  the  observations  of   the  standard  stars
\citep{Land2009},    atmospheric    extinction   and    transformation
coefficients  for the  observing  system were  determined through  the
equations  given  in   \cite{Janes2013}  and  \cite{Aketal2016}.   The
coefficients  for  that particular  night  are  listed in  Table~2  of
\cite{Yonetal2015}.        We        used       Source       Extractor
(SExtractor)\footnote{SExtractor:  Software  for  source  extraction.}
and  PSF   Extractor  (PSFEx)  \citep{BertArn1996,  BertArn2011} 
together with  custom written python  and IDL scripts to  detect and 
measure  and catalog       brightnesses of  the  objects within  the 
cluster's  field  \citep{BertArn1996}.  Aperture  corrections  were 
also applied to the instrumental  magnitudes of the objects identified
in the field.   Instrumental magnitudes  of the objects in the cluster 
field were transformed as described in \cite{Yonetal2015}.

\subsection{Spectroscopy}

The  spectroscopic  observations of  the  stars  in the  direction  of
NGC~225 were performed in two  observing seasons between 2010 and 2011
using the TUG Faint Object Spectrograph and Camera (TFOSC) attached to
the 1.5-m  RTT150 telescope of TUG.  It is possible with  the TFOSC to
obtain  low  resolution ($R\sim  500$)  optical  spectra in  a  single
spectral order  as well  as medium  resolution ($R\sim  5100$) echelle
spectra  of  celestial  objects  in 11  spectral  orders.  During  the
observing  campaign, we  selected the  grism 9  (335-940 nm) to obtain  
the highest possible resolution of  $\sim 5100$. A pinhole with a diameter 
of 100 $\mu$m giving a  spatial resolution of 1$''$.78 were used at
the focal plane of the  telescope. For wavelength calibration, spectra
of  the Fe-Ar lamp  were  taken before  and  after each  program
star observation. We corrected the pixel to pixel variations 
(flat-fielding) using the halogen  lamp spectra taken as  a series in 
the beginning of each observing  night. In  order  to standardize  our 
RV measurements, we selected  two  standard stars  (Vega  and  HD~693) 
and  observed them together with  the program  stars. The reduction 
and analysis  of the spectra were performed with the  IRAF using the 
noao/echelle task.  We obtained a total of 23 spectra  of 21 program 
stars. The observing log is given in Table~2.

\begin{table*}
\setlength{\tabcolsep}{2pt}
\caption{Observing log and some basic information of the program stars
  for which optical spectra were taken in this study. The S/N ratio
  refers to the continuum near H$_{\alpha}$. $V_{r}$ represents the
  radial velocity. Equatorial coordinates, brigtneses and colours of
  the stars located out of our field of view were taken from the
  SIMBAD database and indicated with an asterisk symbol in the first
  column. Star names were taken from \cite{LMM91}. So, LMM is the short 
  for Lattanzi, Massone and Munari.}
{\scriptsize
\begin{tabular}{rcccccccccc}
\hline\hline
ID &  Star   & Other Name & $\alpha_{2000}$ & $\delta_{2000}$ &   $V$  & $B-V$ & SpT&      Date     & S/N &     $V_{r}$       \\
   &         &            &(hh:mm:ss.ss)  &  (dd:mm:ss.ss)  &  (mag) & (mag) &  &(JD-2400000) &     & (km s$^{-1}$)  \\
\hline
$*$01 & LMM~01  &            & 00 41 49.29   &   +61 59 02.71  & 11.320 & 0.500 & A8IV     & 55800.5498  & 100 &    2.2  \\
$*$02 & LMM~33  &            & 00 42 17.90   &   +61 59 35.94  & 12.370 & 0.800 & B4V      & 55798.5729  &  50 &  -79.9  \\
 03   & LMM~83  &            & 00 43 10.88   &   +61 47 19.05  & 10.130 & 0.570 & B8V      & 55486.6114  & 190 &   -6.6  \\
 04   & LMM~88  &  BD+60 86  & 00 43 18.23   &   +61 45 37.56  &  9.713 & 1.668 & K2III    & 55486.5331  & 150 & -129.0  \\
 05   & LMM~92  &            & 00 43 25.63   &   +61 48 51.78  & 11.560 & 0.420 & B9IV     & 55798.3811  &  80 &  -16.0  \\
 06   & LMM~94  &  BD+60 87  & 00 43 26.58   &   +61 45 55.89  & 10.338 & 0.550 & B8IV     & 55797.4427  & 150 &   -4.0  \\
 07   & LMM~95  &            & 00 43 28.88   &   +61 48 04.15  & 10.940 & 0.250 & A0IV     & 55797.5615  & 110 &   -4.3  \\
 08   & LMM~96  &            & 00 43 31.00   &   +61 48 10.29  & 12.172 & 0.551 & A5V      & 55799.3618  & 100 &  -13.4  \\
 09   & LMM~98  &            & 00 43 36.94   &   +61 53 40.23  & 11.982 & 0.347 & A4V      & 55798.5388  &  80 &   -7.1  \\
 10   & LMM~113 & BD+61 157  & 00 43 51.06   &   +61 50 08.44  & 10.630 & 0.280 & B9V      & 55797.4726  & 130 &   -9.5  \\
 11   & LMM~114 &            & 00 43 51.47   &   +61 47 13.54  & 10.890 & 0.230 & B9V      & 55797.5251  & 150 &    0.4  \\
 12   & LMM~115 &  BD+60 91  & 00 43 52.13   &   +61 44 18.04  &  9.873 & 0.961 & G7IV     & 55486.4615  & 160 & -168.2  \\
 13   & LMM~127 &            & 00 44 12.81   &   +61 51 01.88  & 11.420 & 0.290 & A0V      & 55800.5725  & 110 &  -17.4  \\
 14   & LMM~132 &            & 00 44 16.53   &   +61 50 44.01  & 12.342 & 0.436 & A5V      & 55799.4000  & 120 &  -16.1  \\
$**$15& LMM~149 &  BD+60 94  & 00 44 30.68   &   +61 46 49.94  &  9.640 & 0.120 & B8V      & 55797.3182  & 170 &   -9.4  \\
      & LMM~149 &            &               &                 &        &       &          & 55799.5555  & 160 &   -7.6  \\
$**$16& LMM~161 & BD+61 162  & 00 44 40.46   &   +61 54 01.77  &  9.700 & 0.120 & B9V      & 55797.3736  & 170 &   -1.2  \\
$**$17& LMM~163 & BD+61 163  & 00 44 40.82   &   +61 48 43.32  &  9.280 & 0.130 & B7IV-V   & 55797.2775  & 260 &  -16.3  \\
      & LMM~163 &            &               &                 &        &       &          & 55799.5302  & 220 &   -9.8  \\
 18   & LMM~170 & BD+61 164  & 00 44 46.43   &   +61 52 31.52  & 10.004 & 0.149 & B9V      & 55797.6211  & 240 &   -2.4  \\
 19   & LMM~174 &            & 00 44 47.48   &   +61 56 49.42  & 11.608 & 0.386 & A3IV     & 55798.4111  & 110 &  -14.7  \\
 20   & LMM~197 &            & 00 45 08.38   &   +61 56 24.75  & 12.984 & 0.358 & A7III-IV & 55799.4357  & 100 &  -45.8  \\
$*$21 & LMM~269 &            & 00 46 00.67   &   +61 44 14.55  & 11.190 & 0.260 & A5V      & 55799.4990  & 170 &   -5.2  \\
\hline\hline
\end{tabular}
(*): These stars are not located in our field of view.\\
(**): These stars are not included in our photometric catalogue as they are saturated.
}
\end{table*}

Spectral types of the program stars have been determined by means of matching 
the observed spectra with the high-resolution spectra available in the {\sc uves} 
Pop Library \citep{Bagnulo03}. Before starting to match the spectra, the 
resolution difference between two data sets have been corrected. In order to do 
this, we have degraded the high-resolution {\sc uves} spectra by convolving a 
Gaussian $(G(\lambda)=1/(2\pi\sigma^{2}) e^{\lambda^{2}/2\sigma^{2}})$, where 
$\sigma=50$ was found to be best fitting to our spectra. The spectral 
types determined are listed in Table~2. In Fig.~2, we show some spectral 
regions of some selected program stars together with the best matching 
stellar spectra in the {\sc uves} library.

\begin{figure}
\begin{center}
\begin{tabular}{cc}
\resizebox{60mm}{!}{\includegraphics{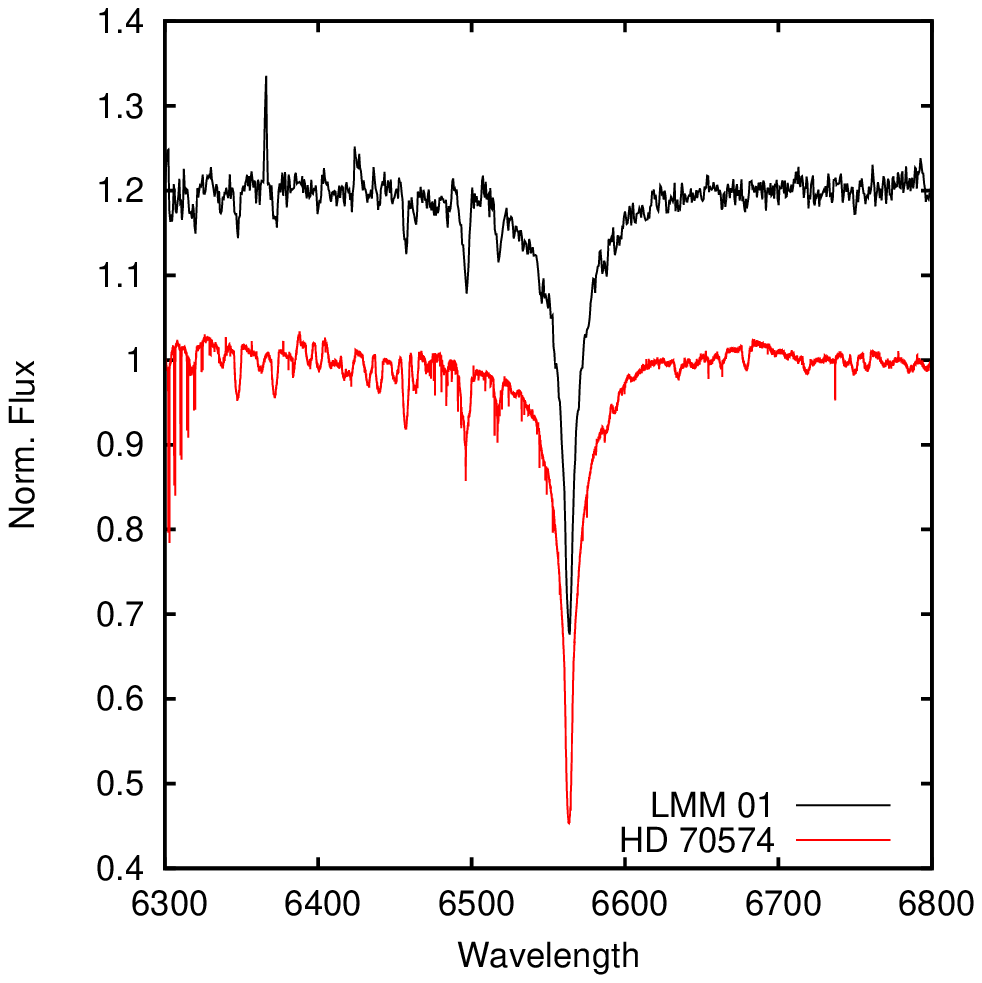}} & \resizebox{60mm}{!}{\includegraphics{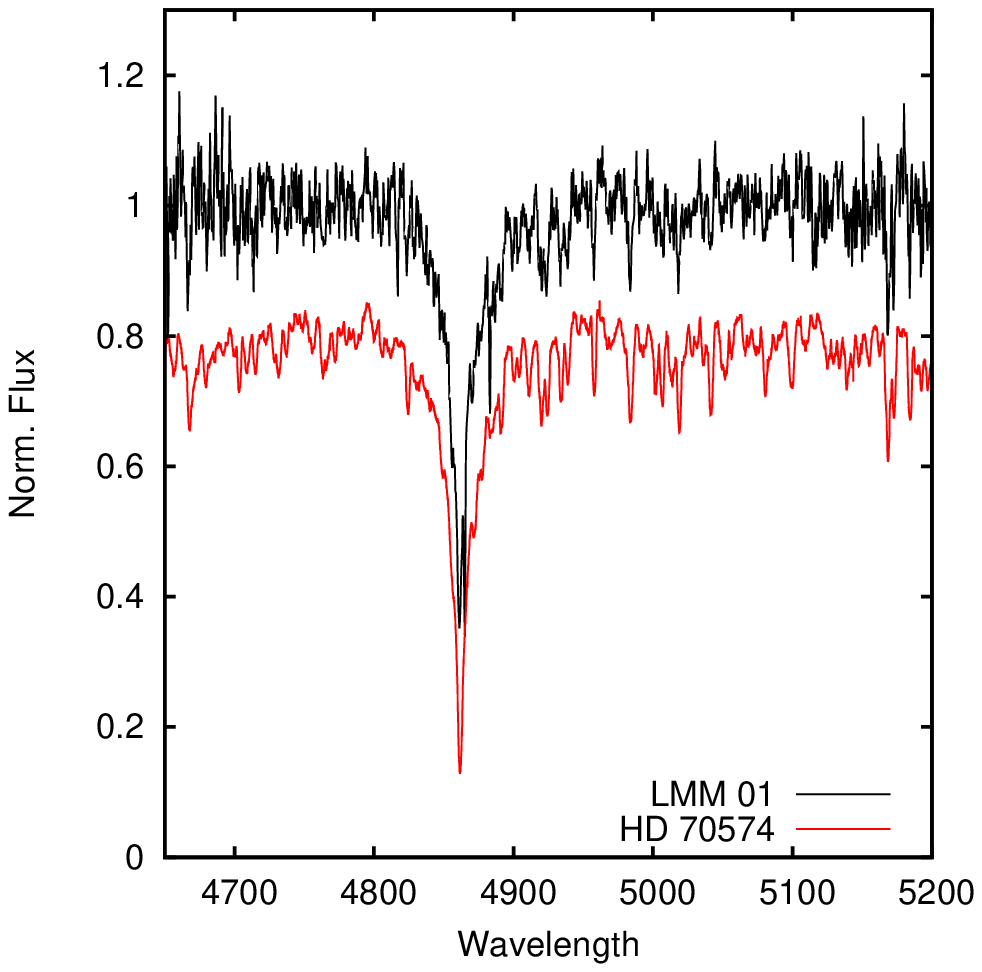}} \\
\vspace{0.5cm}
\end{tabular}
\caption[] {Two sample spectral regions of the selected program star 
and the degraded high-resolution stellar spectra of A8IV-type star HD 70574 
in the {\sc uves} Pop Library.} 
\end{center}
\end{figure}

Optical spectra obtained in this study were also used to measure RVs of the
stars in the direction of NGC~225. Before measuring RVs of the program
stars, possible systematic errors in the RV measurements caused by the
observing  equipments  were investigated  by  means  of measuring  the
positions of the static lines in  the spectra such as the positions of
the telluric lines.  We detected some unexpected Doppler shifts on the
order of several tens of km~s$^{-1}$ due to the motion of the stars in
the  pinhole  at very  small  seeing  conditions and  corrected  these
unexpected shifts by shifting spectra  in the wavelength domain. After
these corrections, the RVs of the program stars were measured by means
of fitting Gaussian functions to the center of spectral lines in
the \textit{splot} task  of IRAF. The uncertainty  in the measurements
is on  the order  of a  few km  s$^{-1}$, which  is determined  by the
standard  deviation  of  the   subsequent  measurements.  However,  we
estimate an uncertainty of $\sim 5-9$~km~s$^{-1}$ for each measurement
due to the  low resolution of spectrograph. The list  of RVs are given
in Table~2.

\section{Data analysis}

\subsection{Identification of stars and the photometric completeness}
We identified 1382  sources in the field of NGC~225  and constructed a
catalogue. The stellarity  index (SI) provided by  SExtractor was used
to  detect   non-stellar  objects,   most  likely  galaxies,   in  our
catalogue. The  objects with SI  smaller than  0.8 were assumed  to be
non-stellar  sources  and  therefore  removed  from  further  analysis
\citep{Andetal2002,Karetal2004}. The resulting catalogue contains 
1019 stars.  Individual stars in the final  photometric  catalogue  are
tabulated  in Table~3.   The columns  of  the table  are organized  as
equatorial coordinates,  apparent magnitude ($V$), colours  ($U-B$, 
$B-V$),  proper  motion   components  ($\mu_{\alpha}\cos  \delta$, 
$\mu_{\delta}$) and  the probability  of membership ($P$).  The proper
motion components of the stars were taken from \cite{Roesetal2010}.

Fig.~3 shows the errors of the measurements in the $V$ band, and $U-B$
and $B-V$  colours as a  function of  the apparent $V$  magnitude.  We
listed the  mean errors in  the selected magnitude ranges  in Table~4.
Table~4 reveals  that the errors  are relatively small for  stars with
$V<17~\rm{mag}$,  while they  increase  exponentially towards  fainter
magnitudes. As expected, the largest  errors for a given $V$ magnitude
were found for the $U-B$ colours of  the stars in the field. For stars
brighter than $V=17$ mag, the mean  photometric errors in the $V$ band
and  $B-V$  colour  index  are  smaller  than  0.012  and  0.021  mag,
respectively. The  mean errors in  the $U-B$ colour index  are smaller
than 0.050 mag for stars brighter than the same $V$ magnitude limit.

\begin{figure}
\begin{center}
\includegraphics[scale=1, angle=0]{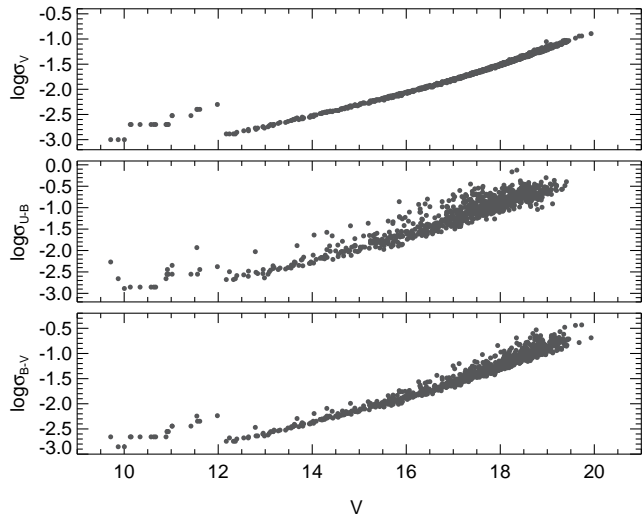}
\caption[] {Colour and magnitude errors of the stars observed in the field 
of the open cluster NGC~225, as a function of $V$ apparent magnitude.} 
\end{center}
\end {figure}

Since  there is  no previously available optical CCD  photometry for  
NGC~225, we have compared our measurements (Fig.~4) with  those of 
\cite{LMM91}, who  give  photographic $V$-band  magnitudes and  $B-V$ 
colour  indices of some stars  in the field of  NGC~225, using 98 
stars  detected both  in their  study and ours. In  Fig.~4, the values 
on  the abscissa refer to  our measurements, while the magnitude or 
colour differences in the ordinates present the differences between 
the two studies. The mean magnitude and colour residuals are calculated 
as $\langle\Delta_{V}\rangle=0.003\pm 0.101$ and $\langle\Delta_{B-V}\rangle=-0.024\pm  0.109$ 
mag from the comparison. This comparison includes only stars 
cross-identified in the two studies and shows generally good agreement 
between them, although the magnitude and colour measurements in 
\cite{LMM91} are based on the photographic photometry for which 
relatively higher mean errors and disagreements with the CCD photometry 
due to different photographic and CCD passbands could be expected.

\begin{figure}
\begin{center}
\includegraphics[trim=1cm 0cm 0cm 0cm, clip=true, scale=0.65]{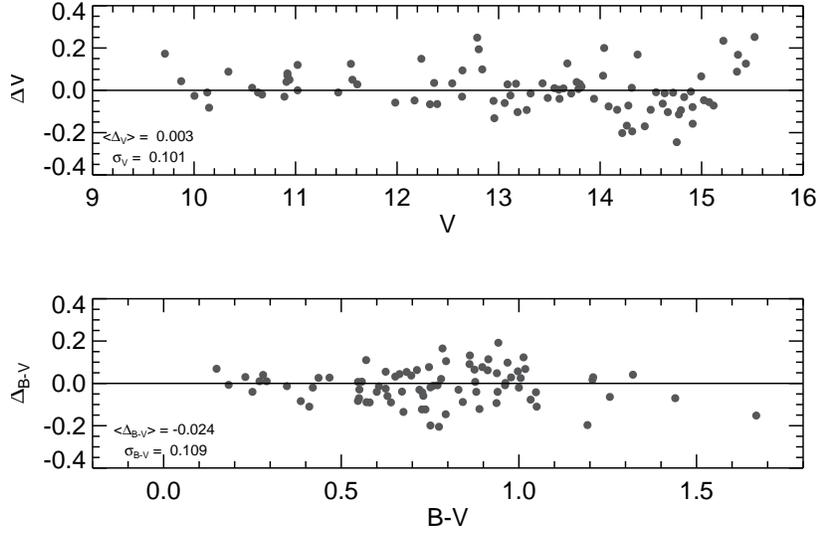}
\caption[] {Comparison of the magnitudes and colours in this study with those of 
\cite{LMM91}. The means and standard deviations of the differences are 
shown in the panels.} 
\end{center}
\end {figure}

\begin{table*}
\setlength{\tabcolsep}{3pt}
\begin{center}
\caption{Photometric and astrometric catalogue for the open cluster NGC~225.
The complete table can be obtained electronically.}
{\scriptsize
\begin{tabular}{ccccccccc}
\hline\hline
 ID & $\alpha_{2000}$ & $\delta_{2000}$ & $V$  &  $U-B$  &  $B-V$  & $\mu_{\alpha}\cos \delta$ & $\mu_{\delta}$ & $P$ \\
   & (hh:mm:ss.ss) &(dd:mm:ss.ss)& (mag) &  (mag) &  (mag)  & (mas yr$^{-1}$)   & (mas yr$^{-1}$) & (\%) \\
\hline
 1 & 00:42:26.59 & 61:51:47.54  & 17.741$\pm$0.025 & 0.826$\pm$0.092 & 1.025$\pm$0.040 & $-$1.3$\pm$3.9  &  $-$5.8$\pm$3.9 & 50 \\
 2 & 00:42:26.65 & 61:55:04.98  & 18.272$\pm$0.039 & 0.313$\pm$0.107 & 1.010$\pm$0.061 & +27.6$\pm$4.0   &  $-$0.1$\pm$4.0 & 0 \\
 3 & 00:42:26.66 & 61:53:14.07  & 17.705$\pm$0.024 & 0.291$\pm$0.054 & 0.777$\pm$0.035 & +18.0$\pm$4.0   &    +3.2$\pm$4.0 & 0 \\
 4 & 00:42:26.73 & 61:50:27.96  & 18.129$\pm$0.034 &       --        & 1.933$\pm$0.088 &   0.0$\pm$3.8   &  $-$1.3$\pm$3.8 & 69 \\
 5 & 00:42:26.75 & 61:46:04.52  & 17.657$\pm$0.024 & 0.326$\pm$0.082 & 1.375$\pm$0.043 & $-$25.2$\pm$5.2 & $-$21.1$\pm$5.2 & 0 \\
 -- &  --  &  -- &  --   &  --  &  --  &  --  &  --  & -- \\
 -- &  --  &  -- &  --   &  --  &  --  &  --  &  --  & -- \\
 -- &  --  &  -- &  --   &  --  &  --  &  --  &  --  & -- \\
\hline\hline
\end{tabular}
}
\end{center}
\end{table*}

\begin{table}
\caption{Mean errors of the photometric measurements for the stars in the field  
of NGC~225. $N$ indicates the number of stars within the $V$ apparent magnitude 
range given in the first column.} 
\begin{center}
\begin{tabular}{ccccc}
\hline\hline
Mag. Range &    $N$ &  $\sigma_V$ & $\sigma_{U-B}$ &  $\sigma_{B-V}$ \\
\hline
$ 9<V\leq11$ & 14  & 0.002 & 0.002 & 0.002 \\
$11<V\leq13$ & 21  & 0.002 & 0.004 & 0.003 \\
$13<V\leq14$ & 25  & 0.002 & 0.005 & 0.003 \\
$14<V\leq15$ & 41  & 0.004 & 0.011 & 0.006 \\
$15<V\leq16$ & 75  & 0.007 & 0.023 & 0.011 \\
$16<V\leq17$ & 116 & 0.012 & 0.050 & 0.021 \\
$17<V\leq18$ & 274 & 0.023 & 0.110 & 0.043 \\
$18<V\leq19$ & 374 & 0.046 & 0.209 & 0.097 \\
$19<V\leq20$ & 79  & 0.077 & 0.286 & 0.163 \\
\hline\hline
\end{tabular}  
\end{center}
\end{table}

\begin{figure}
\begin{center}
\includegraphics[scale=0.70, angle=0]{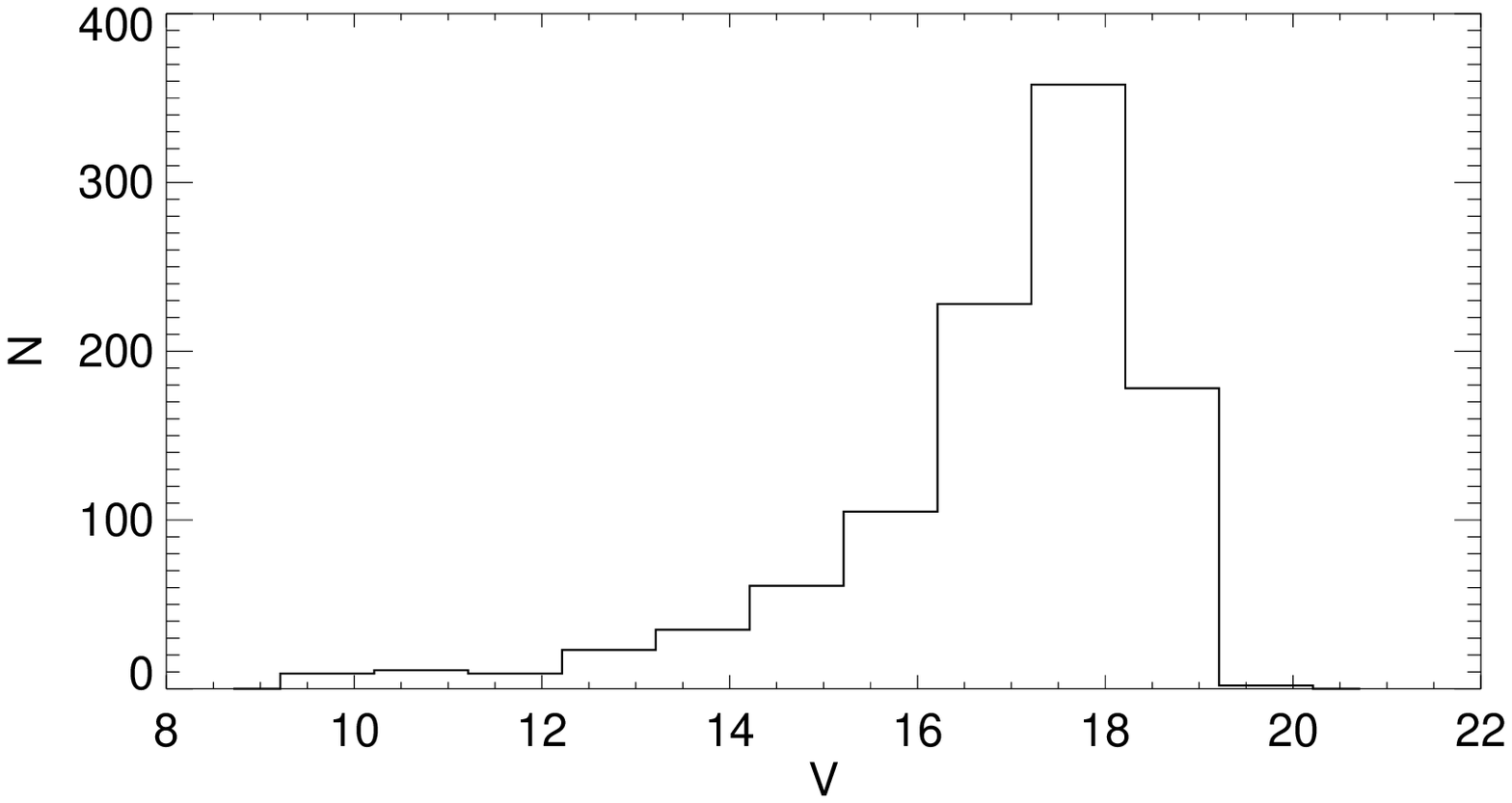}
\caption[] {Histogram of the $V$-band magnitudes measured in the field of 
 the open cluster NGC 225.} 
\end{center}
\end {figure}

We determined the photometric completeness  limit of the data since it
can  be  very important  in  the reliable calculation  of  the
astrophysical parameters of the cluster.  In order to find this limit,
we  constructed  a  histogram  of  $V$ magnitudes  and  showed  it  in
Fig.~5. As the  mode of the distribution of $V$  magnitudes is 18~mag,
we  concluded  that  the  completeness limit  of  the  $V$  magnitudes
corresponds  to   this  magnitude.   This  analysis   shows  that  any
calculation  including only  stars  brighter than  $V=$~18~mag in  our
catalogue will  give reliable results  for the cluster. The number 
of the stars  with  $V\leq$~18~mag in our catalogue  is  566. Note that,
although we found the photometric  completeness limit for the stars in
our  $V$-band  observations,  this   limit  might  not  represent  the
brightnesses of the cluster stars. That  is, most of the cluster 
stars can be much brighter or  fainter than this completeness limit. 
Further analysis, which is given in the following section, is needed 
to conclude if the majority of the cluster's stars are fainter 
or brighter than the completeness limit of our catalogue.

\subsection{Cluster radius and radial stellar surface density}
It  is well-known  that the  structural  parameters of  a cluster  are
calculated by counting the number  of stars in different annuli around
the center of the cluster. First, the stellar density in an area defined  
by a  circle centered  on  the central  coordinates of  the cluster is 
calculated.  From this  central circle, the variation of stellar density  
within annuli with selected widths are calculated. These calculations of 
the stellar density  are used to plot  the stellar  density profile, i.e. 
angular  distance from the centre vs the stellar density. Finally, the 
density profile is fitted with the \cite{King1962} model defined as,

\begin{equation}
\rho(r)=f_{bg}+\frac{f_0}{1+(r/r_{c})^2},
\end{equation}
where  $r$  represents the  radius  of  the  cluster centered  at  the
celestial coordinates: $\alpha_{2000.0}=00^{h}43^{m}39^{s}, 
\delta_{2000.0}=+61^{\circ}46^{'}30^{''}$. $f_{bg}$, $f_0$  and $r_c$ denote
the background  stellar density, the  central stellar density  and the
core radius of  the cluster, respectively. However,  this method might
not give reliable results for a bright and sparse cluster like NGC~225
due to contamination from  background stars. Thus, we followed a
different  approach to  find the  most representative  stellar density
profile for the  cluster with the assumption that  the density profile
of the cluster best fits with the  King model. In order to do this, we
plotted  the density  profiles of  NGC~225 for  the limiting  $V$-band
magnitudes of 15, 16, 17 and 18  and fitted each of them with the King
model. These  density  profiles  and  the best  fits  are  shown  in
Fig.~6. Note that  the number of stars is very  small for the limiting
$V$-band  magnitudes  brighter  than   15  mag.   Using  a  $\chi^{2}$
minimization technique, we found that  the best fitted density profile
is obtained for  the limiting magnitude $V=$ 15.   The $f_{bg}$, $f_0$
and $r_c$  were derived using this  technique, as well.  It  should be
noted    that    the    central    coordinates    of    the    cluster
were assumed to be as given in the WEBDA database. The increase in 
density near 8 arcmin from the center of the cluster (Fig.~6) 
is due to the contamination of faint field stars. Each panels in 
Fig.~6 shows the number of stars brighter than given $V$-magnitude. 
So, as the brightness limit goes to fainter magnitudes, 
fainter and fainter background stars are included in the consecutive 
panels. As a result, we face with a cumulative effect. As it can be 
seen, the effect of these fainter field stars increases to the 
fainter brightness limits. The weakest effect from the fainter 
field stars is found for $V<15$ mag. This increase in the density 
demonstrates that the cluster stars have generally bright apparent 
magnitudes and the cluster has a sparsely distributed structure.  

\begin{figure}
\begin{center}
\includegraphics[trim=0cm 6.0cm 0cm 0cm, clip=true, scale=0.8]{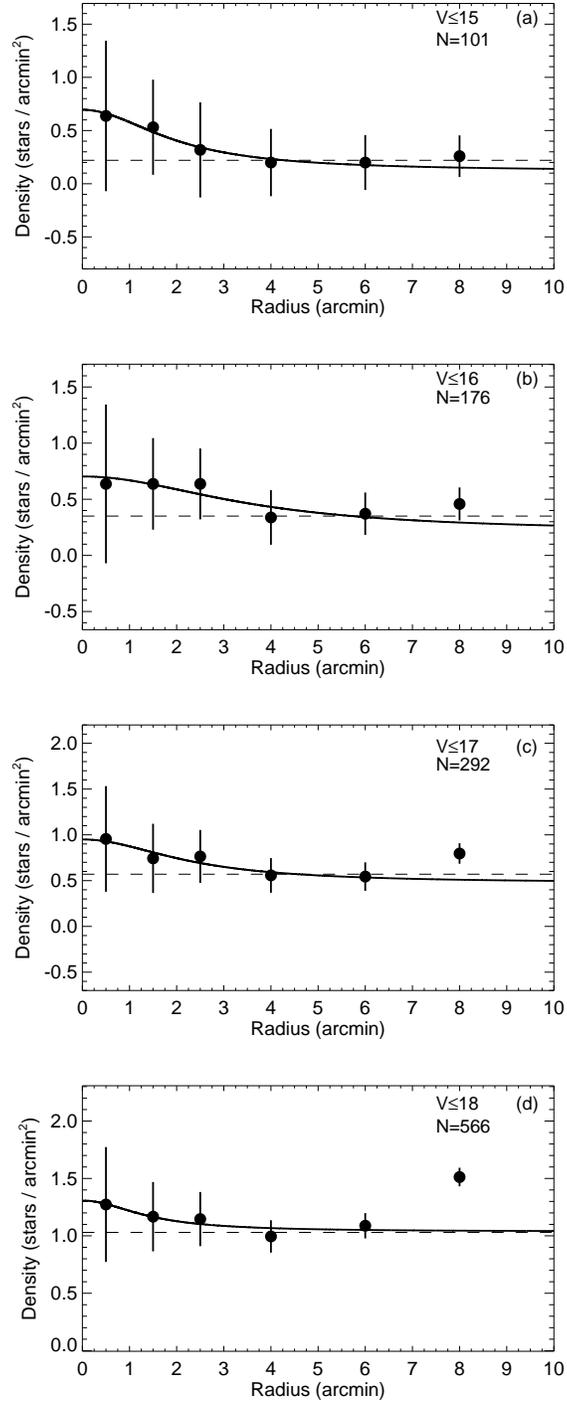}
\caption[] {The stellar density profiles of NGC~225 plotted for the 
limiting $V$-band magnitudes 15 (a), 16 (b), 17 (c) and 18 (d). Solid lines 
represent the best fitted King model for each density profile. Errors were 
determined from sampling statistics: $1/\sqrt{N}$, where $N$ is the number 
of stars used in the density estimation. Dashed lines represent the 
stellar density estimated using the stars with $P\leq50\%$, which 
corresponds to the background level.} 
\end{center}
\end {figure}

From this fit, we derived the  central stellar density and core radius
of  the  cluster, together  with  the  background stellar  density  as
$f_{0}=0.58\pm 0.01$  stars~arcmin$^{-2}$, $r_{c}=1.99\pm 0.08$~arcmin
and $f_{bg}=0.12\pm 0.01$  stars~arcmin$^{-2}$, respectively. The core
radius derived in this study  is smaller than $\sim$5~arcmin estimated
by \cite{LMM91}. An inspection of Fig.~6 by eye shows that the
stellar density of the cluster is  almost equal to that of the stellar
background at about $r=5$ arcmin from  the center of the cluster. 
This analysis also shows that majority of confirmed members in 
the cluster are brighter than $V=$ 15 mag.

\subsection{CMDs and membership probabilities}

The $V$ vs $U-B$ and $V$ vs  $B-V$ CMDs of NGC~225 were constructed to
derive the cluster's astrophysical parameters. The CMDs of the cluster
are shown in Fig.~7. The cluster's main-sequence stars and a few giant 
stars can be distinguished by eye although the cluster is rather sparsely 
populated. The turn-off point with a small  group of bright and blue stars 
in the CMDs of the cluster is located in $10<V<11$ mag.

\begin{figure}
\begin{center}
\includegraphics[scale=0.5, angle=0]{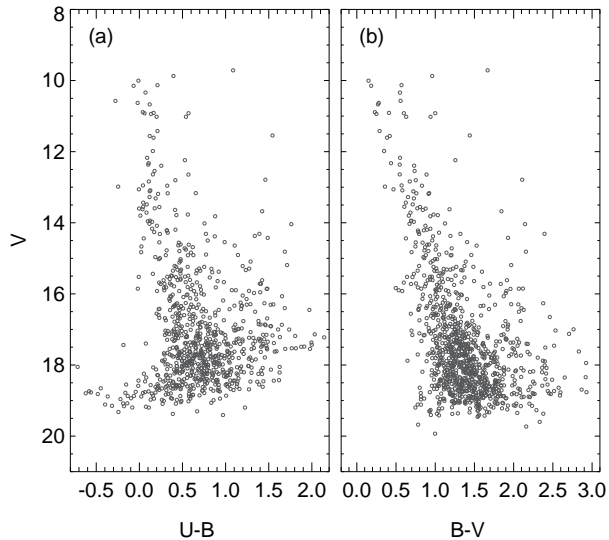}
\caption[] {The CMDs for the cluster NGC 225. (a) $V$ vs $U-B$, 
(b) $V$ vs $B-V$.} 
\end{center}
\end {figure}

One should be careful to determine whether giants and the main-sequence 
stars near the turn-off point of the  CMDs are physical members of the cluster 
or not, before using  them  in the  estimation  of  the  astrophysical 
parameters of the cluster. Identification of the likely members of the
cluster can  also be  useful for the determination  of NGC~225's
main-sequence.   Thus, we  calculated the  probabilities ($P$)  of the
stars  in the  field being  physical members  of the  cluster using  a
method described by \citet{Bala98}.  In this non-parametric method, we
take into account both the errors  of the mean cluster and the stellar
proper   motions,  and   determine  the   cluster  and   field  stars'
distributions empirically  without any  assumption about  their shape.
To  derive  the data  distributions,  we  used the  kernel  estimation
technique  (with a  circular  Gaussian kernel  function).  The  proper
motions of the stars in the CCD field were taken from the PPMXL Catalog of
\cite{Roesetal2010}.  Considering rectangular coordinates of the stars
in  the field,  measured in  two  epochs, first  our observations  and
second  the ones  obtained from  \cite{Roesetal2010}, we  compared our
results with  those of the  algorithm published by  \citet{Java06} and
found  excellent   agreement.   The   histogram  of   the  differences
efficiently  discriminate   the  members  of  the   cluster  from  the
non-members. The  membership probabilities of the  stars identified in
the field of NGC~225 are listed in the last column of Table~3.

\begin{figure}
\begin{center}
\includegraphics[trim=0.5cm 0cm 0cm 0cm, clip=true, scale=0.7]{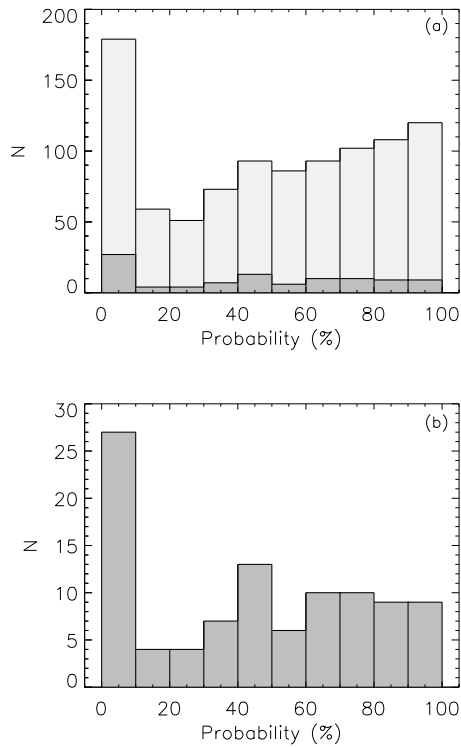}
\caption[] {(a) Histogram of the membership probabilities of the stars 
in our catalogue for the magnitude ranges $10\leq V \leq 19$ (light shaded 
bars) and $10\leq V \leq 15$ mag (shaded bars). (b) Histogram of the 
membership probabilities of the stars in the magnitude range $10\leq V \leq 15$ 
mag is shown in a larger scale.} 
\end{center}
\end {figure}

In order to determine the most likely members of the cluster, we first
plotted  the  histogram  of  membership probabilities  for  the
$V$-band  magnitudes  of  1019  stars   identified  in  the  field  of
NGC~225. We also  overplotted this histogram with  the one constructed
for the stars, whose apparent magnitudes  are $10\leq V \leq 15$ which
is shown  with grey  bars in  Fig.~8a.  The  above analysis  about the
structural parameters show  that the majority of  the cluster's member
stars are in this magnitude  range.  In addition, the magnitude $V=10$
roughly corresponds to  the turn-off point of the  cluster. The number
of stars  with $10\leq V  \leq 15$ mag in  our catalogue is  only 101.
Their  histogram  is  presented  in  Fig.~8b.   Median  value  of  the
membership probabilities  of these stars corresponds  to $\sim$50$\%$.
Thus, we concluded that the stars with $P\geq 50\%$ are likely members
of the cluster. 

The  zero age  main-sequence (ZAMS)  of \cite{Sungetal2013}  for solar
metallicity  could be  used  to  determine the  most  likely members  of
NGC~225 on the main-sequence.  Thus, we fitted this ZAMS to the $V$ vs
$B-V$ CMD  of the cluster  for $10\leq V \leq  15$ mag using  only the
stars  with $P\geq  50\%$.  By  shifting the  fitted main-sequence  to
brighter $V$ magnitudes  by 0.75 mag (see Fig.~9), a  band like region
in the  $V$ vs $B-V$  CMD was obtained to  cover the binary  stars, as
well. Hence, we  assumed that all stars with  a membership probability
$P\geq 50\%$  and located  within this band-like  region are  the most
likely main-sequence members of NGC~225, resulting in 28 stars. A visual
inspection  in Fig.~9 indicates that  there are  stars with  $P\geq
50\%$ that have already  left the  ZAMS as shown by the red dots 
outside the main-sequence band. We assume  that they are also 
likely members of the cluster. With  this procedure, we identified 
31 stars for further analyses.

\begin{figure}
\begin{center}
\includegraphics[trim=1.4cm 0cm 0cm 0cm, clip=true, scale=0.55]{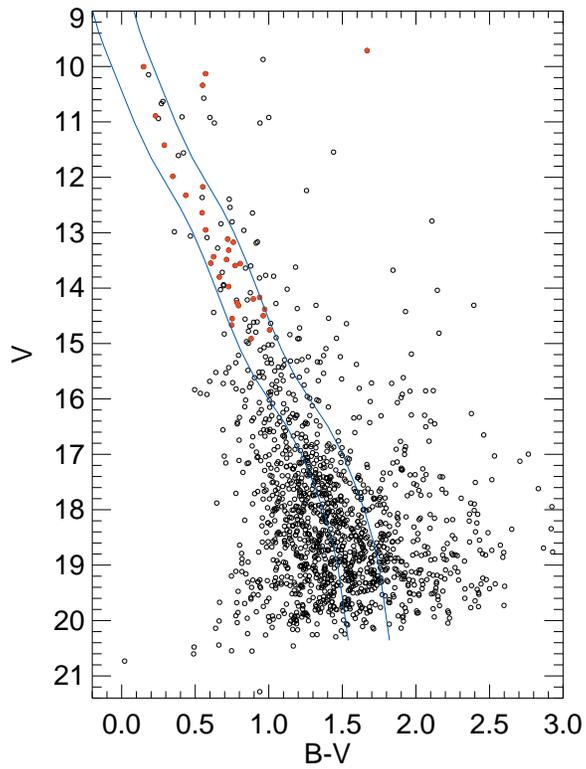}
\caption[] {$V$ vs $B-V$ CMD of NGC~225 constructed using all stars in our catalogue. 
 Solid lines represent the ZAMS of \cite{Sungetal2013} and the one shifted by an 
 amount of 0.75 mag to the bright $V$ magnitudes. Red dots indicate the most
 probable cluster stars that are identified using a procedure explained in the text.} 
\end{center}
\end {figure}

\section{Determination of the astrophysical parameters of NGC~225}

\subsection{The reddening}

Interstellar reddening affects the TCDs and  CMDs, from which
the remaining astrophysical parameters  will be determined.  Thus, the
colour  excesses   $E(U-B)$  and  $E(B-V)$  must   be  inferred 
first. For  the determination  of these colour  excesses, we  used the
most  probable 28  main-sequence  stars in  the $10  \leq  V \leq  15$
apparent  magnitude  range,  which  were  selected  according  to  the
procedure in Section  3.3.  The positions of these stars  in the $U-B$
vs $B-V$ TCD were compared with the ZAMS of \cite{Sungetal2013} with a
solar metallicity.  In order to do this, the de-reddened main-sequence
curve  of  \cite{Sungetal2013} was  shifted  with  steps of  0.001~mag
within the  range $0\leq E(B-V)\leq  0.60$~mag  until the best  fit is
obtained with  the $U-B$ vs  $B-V$ TCD of  NGC~225.  The shift  in the
$U-B$  axis   was  calculated  by  adopting   the  following  equation
\citep{Cox2000}:

\begin{equation}
E(U-B) = E(B-V) \times [0.72 + 0.05 \times E(B-V)].
\end{equation}

We  show the  $U-B$ vs  $B-V$  TCD of  NGC~225 for  the most  probable
main-sequence stars of the cluster in  Fig.~10. The goodness of the fit
was determined  by adopting the  minimum $\chi^2$ method.   Using this
method, we estimated the  following colour excesses: $E(U-B)=0.110 \pm
0.034$ and $E(B-V)=0.151 \pm 0.047$ mag.  The errors indicate the $\pm
1\sigma$ deviations.

\begin{figure}
\begin{center}
\includegraphics[scale=0.7, angle=0]{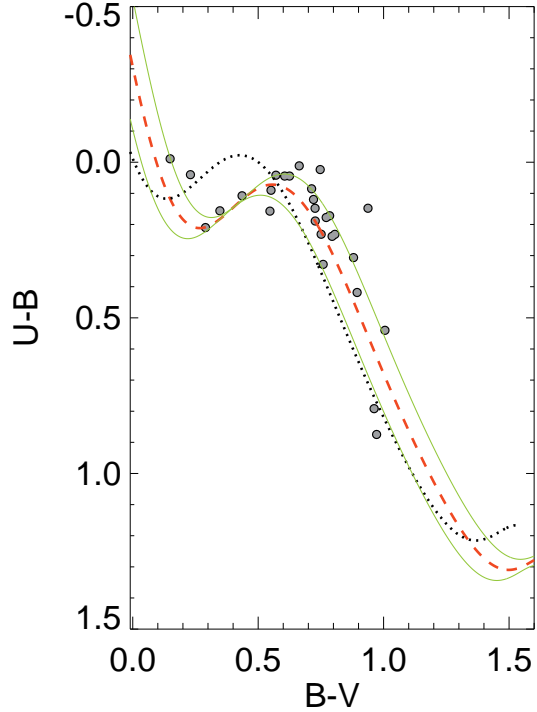}
\caption[] {$U-B$ vs $B-V$ TCD for the main-sequence stars with 
$10\leq V \leq 15$ mag in NGC~225. The reddened and de-reddened main-sequence 
curves \citep{Sungetal2013} fitted to the cluster stars are represented with 
red dashed and black dotted lines, respectively. Green lines represent 
$\pm 1\sigma$ deviations. The number of stars is 28.} 
\end{center}
\end {figure}

\subsection{Photometric metallicity of NGC~225}

The metallicity of NGC~225 has not been determined in previous studies.
We  used the  method described  in \cite{Karaali03, Karaali05, Karetal2011} to measure  the
photometric metallicity.  The procedure in  this method uses  F-G type
main-sequence stars  of the cluster. Thus,  we selected 9 of  28 stars
with  colours $0.3\leq  (B-V)_0\leq  0.6$ mag  corresponding to  F0-G0
spectral type main-sequence stars \citep{Cox2000}.

\begin{figure}
\begin{center}
\includegraphics[scale=0.7, angle=0]{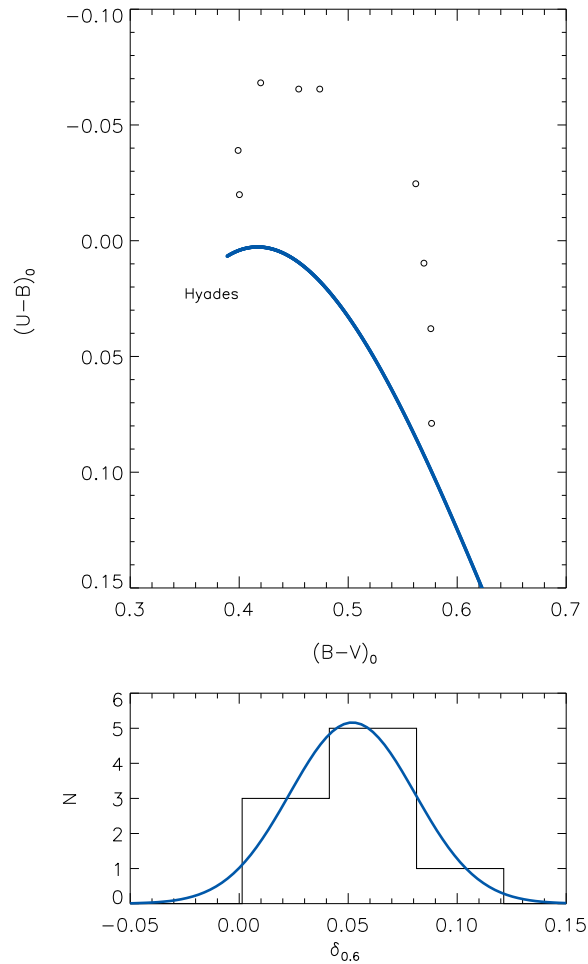}
\caption[] {The $(U-B)_0$ vs $(B-V)_0$ TCD (upper panel) and the histogram 
(lower panel) for the normalized UV-excesses for nine main-sequence stars used 
for the metallicity estimation of NGC 225. The solid lines in the upper and 
lower panels represent the main-sequence of Hyades cluster and the Gaussian 
fit of the histogram, respectively.} 
\end{center}
\end {figure}
 
The difference  between a  star's de-reddened $(U-B)_0$  colour indice
and the  one corresponding to the  members of the Hyades  cluster with
the  same de-reddened  $(B-V)_0$  colour index  is  the normalized  UV
excess         of         the        star         in         question,
i.e.  $\delta=(U-B)_{0,H}-(U-B)_{0,S}$. Here,  the subscripts  $H$ and
$S$ refer  to Hyades and star,  respectively. In order to  utilize the
method described  in \cite{Karetal2011}, we calculated  the normalized 
UV excesses of the nine stars selected as 
described above and normalized their $\delta$ differences to the 
UV-excess at $(B-V)_{0}=0.6$ mag, i.e. $\delta_{0.6}$. The $(U-B)_0$ vs $(B-V)_0$
TCD and the  histogram of the normalized  UV excesses ($\delta_{0.6}$)
of the selected  nine main-sequence stars of NGC~225  are presented in
Fig.~11. By  fitting a Gaussian  to this histogram, we  calculated the
normalized UV  excess of the cluster  as $\delta_{0.6}=0.051\pm 0.002$
mag.  Here, the uncertainty is given as the statistical uncertainty of
the  peak  of  the  Gaussian.   Then,  we  estimated  the  metallicity
($[Fe/H]$) of  the cluster by  evaluating this Gaussian peak  value in
the following equation of \cite{Karetal2011}:

\begin{eqnarray}
[Fe/H]=-14.316(\pm 1.919)\delta_{0.6}^2-3.557(\pm 0.285)\delta_{0.6}\\ \nonumber 
+ 0.105(\pm 0.039). 
\end{eqnarray}

The metallicity corresponding to the peak value for the $\delta_{0.6}$
distribution was calculated  as $[Fe/H]= -0.11\pm 0.01$  dex. In order
to transform the $[Fe/H]$ metallicity  obtained from the photometry to
the   mass   fraction   $Z$,   the   following   relation   was   used
\citep{Mowlavietal2012}:

\begin{equation}
Z=\frac{0.013}{0.04+10^{-[Fe/H]}}.
\end{equation}
Here, $Z$  is the mass fraction  of all elements heavier  than helium,
which  is  used  to  estimate  the  theoretical  stellar  evolutionary
isochrones.  Hence,  we  calculated  $Z=0.009$  from  the  metallicity
($[Fe/H]=-0.11$ dex)  obtained from  the photometry.  This metallicity
will be used for the further analysis in this study.

\begin{table*}
\setlength{\tabcolsep}{3pt}
\caption{Colour excesses, metallicities ($Z$), distance moduli ($\mu$), distances ($d$) 
and ages ($t$) estimated using two CMDs of NGC~225.}
\begin{center}
\begin{tabular}{lccccc}
\hline
CMD          & Colour Excess & $Z$   & $\mu_{V}$ &  $d$    &  $t$    \\
             &      (mag)    &       &   (mag)   &  (pc)   & (Myr)    \\
\hline
$V$ vs $U-B$ & $E(U-B)=0.110\pm0.034$ & $0.009$ & $9.30\pm 0.07$ &  $585\pm 20$ & $900\pm100$ \\
$V$ vs $B-V$ & $E(B-V)=0.151\pm0.047$ & $0.009$ & $9.30\pm 0.07$ &  $585\pm 20$ & $900\pm100$ \\
\hline
\end{tabular}
\end{center}
\end{table*}

\subsection{Distance modulus and age of NGC~225}

We fitted the $V$  vs $U-B$ and $V$ vs $B-V$ CMDs  of the cluster with
the theoretical  isochrones provided  by the PARSEC  synthetic stellar
library \citep{Bresetal2012}, which was recently updated \citep[PARSEC
  version 1.2S,][]{Tangetal2014,Chenetal2014}  to derive  the distance
modula and  age of  NGC~225 simultaneously.   Since the  reddening and
metallicity of the  cluster were above determined  using its $(U-B)_0$
vs $(B-V)_0$ TCD  and the normalized ultraviolet (UV)  excesses of the
cluster  members,  respectively,  we  kept  the  metallicity  and  the
reddening as constants  during the fitting process.   As already noted
in \cite{Yonetal2015} and  \cite{Bosetal2015}, the measured reddening,
metallicity,  and  therefore  the  age  values  can  suffer  from  the
degeneracies  between   the  parameters  when  these   parameters  are
determined  simultaneously.  Thus,  uncertainty  of the  age and
reddening  can   be  higher   than  expected  from   the  simultaneous
solutions. To overcome this problem, we determined the metallicity and
reddening of the cluster using independent and reliable methods
and kept  these parameters  as constants in  the analysis.   We expect
that the degeneracy/indeterminacy of the parameters determined in this
study will  be less than that  in the statistical solutions  with four
free  astrophysical  parameters  (i.e. the reddening, distance modulus, 
metallicity and age).

The estimated  astrophysical parameters  of NGC~225 obtained  from the
best fits to  the CMDs are given in Table~5.  Errors of the parameters
were  derived  by  visually  shifting the  theoretical  isochrones  to
include  all  the  main-sequence  stars  in the  CMDs.  The  best  fit
theoretical isochrones  for $Z=0.009$ and $t=900~\rm{Myr}$  in the $V$
vs $B-V$ and $V$ vs $U-B$ CMDs are overplotted in Fig.~12.

\begin{figure*}
\begin{center}
\includegraphics[scale=0.75, angle=0]{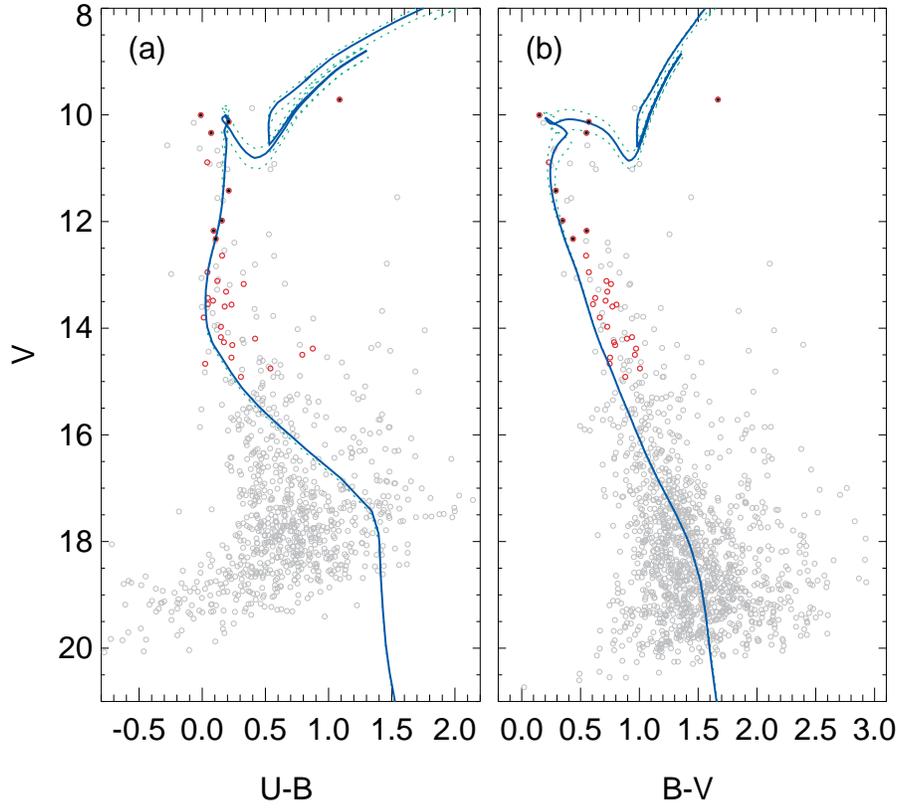}
\caption[] {\small $V$ vs $U-B$ (a) and $V$ vs $B-V$ (b) CMDs for the 
stars in the field of NGC~225. The most probable members of the cluster 
are indicated with red circles. These stars are fitted to the isochrone 
determined in this study (blue line). The green dots indicate the 
isochrones with estimated age plus/minus its error. Stars indicated with 
black dotes are the ones with spectra from which RVs were measured in 
this work.}
\end{center}
\end{figure*}

\begin{table*}
\setlength{\tabcolsep}{2pt}
\begin{center}
\caption{The data for stars used in the calculation of the galactic orbit and 
space velocity components of NGC~225. LMM 88 is a spectroscopic binary star which was 
not used in these calculations. $V_{r}$ denotes the radial velocity measured from the 
spectra in this study. Star names were taken from the SIMBAD Database.}
{\scriptsize
\begin{tabular}{ccccccccc}
\hline\hline
Star & $\alpha_{2000}$ & $\delta_{2000}$ &  $V$  & $B-V$ & $\mu_{\alpha}\cos \delta$ & $\mu_{\delta}$  &    $V_{r}$    &  $P$  \\
     & (hh:mm:ss.ss)   & (dd:mm:ss.ss)   & (mag) & (mag) &     (mas yr$^{-1}$)       & (mas yr$^{-1}$) & (km s$^{-1}$) &  (\%) \\
\hline
LMM~83  & 00 43 10.86 & +61 47 18.96 & $10.130\pm 0.002$  &  $0.570\pm 0.002$  & $-2.7\pm 2.0$  & $-1.8\pm 2.0$ &   -6.6 & 64 \\
LMM~94  & 00 43 26.59 & +61 45 55.82 & $10.338\pm 0.002$  &  $0.550\pm 0.002$  & $-4.0\pm 2.0$  & $+0.6\pm 2.0$  &   -4.0 & 96 \\
LMM~96  & 00 43 30.99 & +61 48 10.25 & $12.172\pm 0.001$  &  $0.551\pm 0.002$  & $-4.8\pm 2.0$  & $-1.0\pm 2.0$ &  -13.4 & 92 \\
LMM~98  & 00 43 36.93 & +61 53 40.08 & $11.982\pm 0.005$  &  $0.347\pm 0.006$  & $-6.2\pm 2.0$  & $+2.4\pm 2.0$  &   -7.1 & 67 \\
LMM~114 & 00 43 51.48 & +61 47 13.43 & $10.890\pm 0.002$  &  $0.230\pm 0.001$  & $-3.8\pm 2.0$  & $-2.2\pm 2.0$ &    0.4 & 69 \\
LMM~127 & 00 44 12.80 & +61 51 01.89 & $11.420\pm 0.003$  &  $0.290\pm 0.004$  & $-5.9\pm 2.0$  & $+1.3\pm 2.0$  &  -17.4 & 86 \\
LMM~132 & 00 44 16.55 & +61 50 44.05 & $12.324\pm 0.001$  &  $0.436\pm 0.002$  & $-6.8\pm 2.0$  & $-0.3\pm 2.0$ &  -16.1 & 77 \\
LMM~170 & 00 44 46.42 & +61 52 31.44 & $10.004\pm 0.001$  &  $0.149\pm 0.001$  & $-5.4\pm 2.0$  & $+1.7\pm 2.0$  &   -2.4 & 86 \\
\hline\hline
\end{tabular}
}
\end{center}
\end{table*}

\subsection{Galactic orbit of the cluster}
The  procedure of estimating the galactic orbital parameters of an object  
is described in \cite{Dinetal1999}, \cite{Cosetal12} and \cite{Bil12}. 
To  estimate galactic orbital parameters of NGC~225, we first performed a 
test-particle integration in a Milky Way potential which consists of a
logarithmic halo, a Miyamoto-Nagai potential to represent the galactic
disc and a Hernquist potential to model the bulge.

We  calculated galactic  orbits  of  the eight  cluster  stars with  a
membership probability larger than 50$\%$  for which RV data are given
in Table~6. Mean  values of the RV, proper motion  components of these
stars  and  the distance  of  the  cluster  were  taken as  the  input
parameters  for  the  cluster's  galactic  orbit  estimation:  $V_{r}=
-8.3\pm 5.0$ km  s$^{-1}$, $\mu_{\alpha}\cos{\delta}=-4.95\pm 2.0$ and
$\mu_{\delta}=0.09\pm  2.0$  mas  yr$^{-1}$,  and  $d=585\pm  20$  pc,
respectively.  \cite{Meretal08}  measured the RV of  LMM 88 (BD+60~86)
as  $-33.86\pm 0.76$  km s$^{-1}$,  which is  very different  than the
value  in  Table  6.   Therefore,  this star  was  excluded  from  the
calculation of  the mean  RV of  the cluster, since  it is  probably a
spectroscopic  binary. The  proper motion  components were  taken from
\cite{Roesetal2010},  while the  RVs of  the cluster  stars and  their
distances were found in  this study.   We determined  galactic orbits  
of the  stars within  an  integration  time  of  3~Gyr  in  steps  of  
2~Myr. This integration  time corresponds  to  minimum 12  revolutions 
around  the galactic  center  so  that  the averaged  orbital  
parameters  can  be determined reliably.

In order  to determine the galactic  orbit of the cluster,  we adopted
means of  the orbital parameters  found for  the stars as  the orbital
parameters  of  the cluster.   Fig.~13  shows  the representations  of
galactic  orbits calculated  for  the cluster  stars  and the  cluster
itself in  the $X-Y$  and $X-Z$  planes.  Here, $X$,  $Y$ and  $Z$ are
heliocentric  galactic  coordinates   directed  towards  the  galactic
centre, galactic  rotation and the north  galactic pole, respectively.
We  obtained  the  cluster's apogalactic  ($R_{a}$)  and  perigalactic
($R_{p}$)  distances   as  $9.37\pm  0.15$  and   $8.22\pm  0.12$~kpc,
respectively. The maximum vertical distance from the galactic plane is
calculated as $Z_{max}=  90\pm 70$~pc.  We used  the following formula
in the determination of the  eccentricity projected on to the galactic
plane: $e=(R_{a}-R_{p})/(R_{a}+R_{p})$. The  eccentricity of the orbit
was calculated as $e=0.07\pm 0.01$.  This value shows that the cluster
is orbiting  the Galaxy with  a period  of $P_{orb}= 255\pm  5$~Myr on 
nearly circular orbit as expected for the objects in the solar 
neighbourhood.

We also computed  the galactic space velocity components  of the stars
in  Table~6  with  respect  to   the  Sun  using  the  algorithms  and
transformation matrices of \cite{JS87}.  The input data in Table~6 are
in  the form  adopted  for the  epoch  of J2000  as  described in  the
International Celestial Reference System (ICRS) of {\it Hipparcos} and Tycho
Catalogues \citep{ESA97}. The calculated  space velocity components are
$(U,V,W)=(16.02\pm  5.41, 0.25\pm  5.17, 0.82\pm  5.55)$ km  s$^{-1}$.
The uncertainties of these components were computed by propagating the
uncertainties of the input data (proper motions, distance and RV) with
an  algorithm also  by \cite{JS87}.   We applied  corrections for  the
differential  galactic rotation  to the  space velocity  components as
described  in  \cite{MB81}.  We  also  corrected  the  galactic  space
velocity components for the Local Standard of Rest (LSR) by adding the
space velocity  of the  Sun to  the space  velocity components  of the
stars. The adopted space velocity of the Sun is $(U,V,W)=(8.50, 13.38,
6.49)$ km  s$^{-1}$ \citep{Coetal11}.  We adopted  means of  the space
velocity  components  found  for  the  stars  as  the  space  velocity
components  of  the cluster.  Finally,  the  corrected space  velocity
components of NGC~225 were  found as $(U,V,W)=(11.03\pm 5.41, 14.37\pm
5.17,  7.31\pm  5.55)$  km  s$^{-1}$. We  estimated  the  total  space
velocity of  the cluster as $S_{tot}=19.53\pm9.32$  km s$^{-1}$, which
is in agreement  with the one suggested for the  young thin-disc stars 
and young stellar clusters \citep{Leg92}.

\begin{figure}
\begin{center}
\includegraphics[trim=1cm 1cm 0cm 1cm, clip=true, scale=0.42]{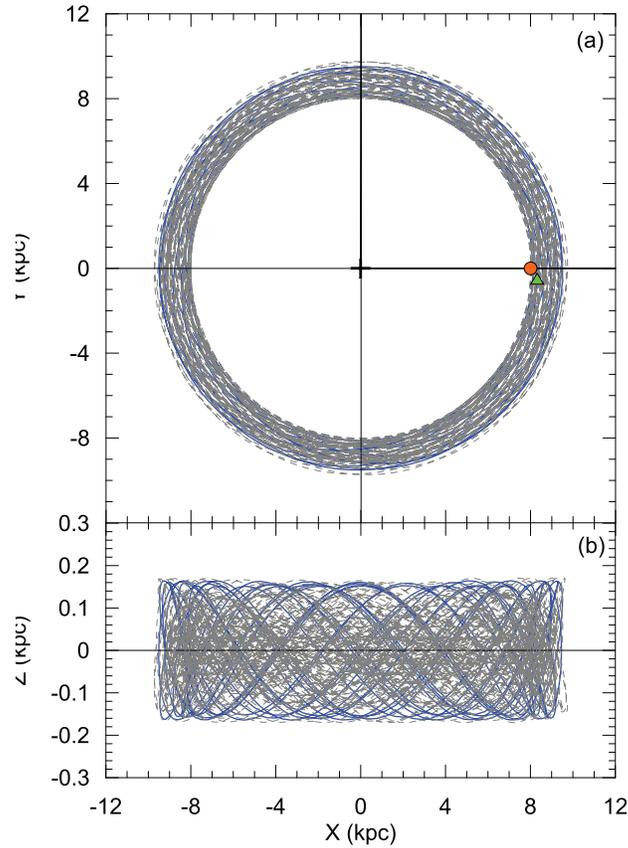}
\caption[] {\small The galactic orbital motions (grey dashed lines) of 
eight cluster stars with P $\geq$ 50$\%$, for which RVs are available, 
in the $X-Y$ (a) and $X-Z$ (b) planes. The cluster's mean orbit is 
indicated with a blue line. The black plus, red circle and green triangle 
symbols in panel (a) represent the galactic centre, and current locations 
of the sun and NGC 225, respectively.}
\end{center}
\end{figure}

\section{Discussion}

In this paper, we present the  first CCD $UBV$ photometry for the open
cluster  NGC~225.  From  these  data,  we  determined  structural  and
astrophysical parameters of  the cluster. We also  calculated the mean
RV of the  cluster from the spectroscopic observations  of the cluster
members.  Using the  proper  motions of  the stars  in  the field,  we
estimated probability of the stars in the cluster field being physical
members  of  the  cluster.  We  also  calculated  the  space  velocity
components and the parameters of the galactic orbit of NGC~225.

Independent methods  developed for the determination  of the reddening
and metallicity were  used in this study to reduce  the number of free
parameters  in  the  simultaneous  solutions,  where  the  theoretical
stellar evolutionary isochrones are fitted to the observed CMDs, since
the   astrophysical  parameters   of   a  cluster   suffer  from   the
reddening-age degeneracy when all of them (the reddening, metallicity,
distance modulus  and the age)  are simultaneously determined  by this
fitting  process   \citep[cf.][]{Anders2004,  Kingetal2005,  Brid2008,
  deMeule2013}.  Thus, we  inferred the reddening of  the open cluster
NGC~225  from  its  $U-B$  vs  $B-V$  TCD,  while  we  determined  the
metallicity of the cluster utilizing F0-G0 spectral type main-sequence
stars  \citep{Cox2000}  via  a   metallicity  calibration  defined  by
\cite{Karetal2011}.   The reddening  and  metallicity  of the  cluster
obtained from the $UBV$ photometric  data are $E(B-V)= 0.151\pm 0.047$
mag and  $[Fe/H]= -0.11\pm 0.01$ dex  ($Z=0.009$), respectively.  This
is  the first  determination of  the metallicity  for NGC~225.   While
keeping  the  reddening  and  metallicity constant,  we  inferred  the
distance modulus and age of the  cluster as $(m-M)= 9.30\pm 0.07$ mag and
$t=  900\pm   100$~Myr,  respectively,  by  fitting   the  theoretical
isochrones to the  observed CMDs. From the  estimated distance modulus
and the  colour excess, we find  the distance of NGC~225  as $d=585\pm
20$~pc, which is  in agreement with that found  by \cite{LMM91} within
errors. A comparison  of the reddening values with  those estimated in
previous studies shows  that the colour excess obtained  in this study
is smaller  than $E(B-V)=  0.29$~mag given by  \cite{Johetal61} and
\cite{Svo62}, while it  is in agreement roughly  with $E(B-V)= 0.25\pm
0.08$~mag  found  by \cite{LMM91} within  errors. The distance modulus
obtained in this study $(m-M)= 9.30\pm0.07$ mag is in agreement with the
previous estimates of $(m-M)=9.0$ and $(m-M)=9.1$~mag \citep{Svo62,HoApp65} 
within errors, as well.

We compared the spectral types and membership probabilities estimated 
in our study with those in \cite{LMM91}. The comparison is shown in Table 7.     
Stars in the Table 7 are indicated as the most probable members of the 
cluster by \cite{LMM91}. An inspection by eye shows that spectral types 
and membership probabilites in the two studies are generally in agreement. 
Differences between the probabilites are due to the proper motion accuracies 
of the catalogues used in the two studies. Since we have taken the proper motion 
values from \cite{Roesetal2010}, proper motion errors in our study are smaller 
than those in \cite{LMM91}, indicating that the membership estimated in our 
study are more reliable.

As  for the  age of  NGC~225, \cite{LMM91}  estimated the  age of  the
cluster as $t= 120$ Myr, while \cite{SMK06} argued that the age should
be $t=0.5-10$ Myr  based on pre-main sequence  isochrones. However, we
found age  of the  cluster to  be much older than both determinations as 
$t= 900\pm  100$~Myr by  fitting the theoretical  isochrones  provided  
by  the  PARSEC  synthetic  stellar
library \citep{Bresetal2012}, which was recently updated \citep[PARSEC
  version 1.2S,][]{Tangetal2014,Chenetal2014}, to  the observed $V$ vs
$U-B$ and  $V$ vs $B-V$  CMDs in  this study. 
\cite{SMK06}  show  that  there  are  eight  stars  with  H$_{\alpha}$
emission located around the cluster,  of which two are probable Herbig
Be  stars  (LKH$_{\alpha}$~201  and  BD~+61~154)  located  within  the
cluster's field.  According to  \cite{SMK06}, these two stars indicate
that the age  of NGC~225 is 1.5$-$3.0~Myr, if they  are members of the
cluster. The re-reduced {\it Hipparcos} parallax of BD~+61~154 is 
$\pi   =4.96\pm  1.61$   mas  \citep{vLe07}. This value compares to the 
cluster's parallax ($1.71\pm 0.06$ mas) only within 2$\sigma$ error level. 
It should be noted that the GAIA observations will reveal if this star is 
a member of the cluster. \cite{SMK06}  calculated  the  distance  of
LKH$_{\alpha}$~201 as $d=1500$  pc and showed that the  star is behind
the cluster.   So, LKH$_{\alpha}$~201 is  not a member of  NGC~225, as
well. Thus, we  conclude that presence of these stars  in the field of
the cluster is just a coincidence. 

In addition, LKH$_{\alpha}$~200 was also used by \cite{SMK06} to 
show that the cluster is very young. However, the membership 
probabilities of LKH$_{\alpha}$~200 and LKH$_{\alpha}$~201 are 
estimated to be 0\% in our study as well as in \cite{LMM91} (see Table 7). 
Thus, LKH$_{\alpha}$~200 can not be used in the estimation of the 
cluster's age, as well.

\begin{landscape}
\begin{table*}
\setlength{\tabcolsep}{2pt}
{\scriptsize
\begin{center}
\caption{Comparison of the photometry, spectral types and membership 
probabilities given in \cite{LMM91} and this study.}
\begin{tabular}{ccccccccc|cccccl}
\hline\hline
 & & & \multicolumn{6}{c}{\cite{LMM91}}&\multicolumn{6}{c}{This Study}\\
\cline{4-15}\\
ID & $\alpha_{2000}$ & $\delta_{2000}$  & Star       & Star   & SpT & $V$ & $B-V$ & $P$  & SpT & $V$ & $U-B$ & $B-V$ & $P$ & Remark\\
   &(hh:mm:ss.ss)&(dd:mm:ss.ss)&        &            &          &(mag)& (mag) & (\%) &     &(mag)& (mag) & (mag) &  (\%) & \\   
\hline\hline
01 & 00:42:12.40 & 61:38:16.49 & LMM 29 & ---        & F8  & 13.44 & 0.67 & 58 & --- & --- & --- & --- & --- & out of FoV \\
02 & 00:42:17.90 & 61:59:35.94 & LMM 33 & ---        & F   & 12.55 & 0.57 & 93 & B4V & --- & --- & --- & --- & out of FoV \\
03 & 00:42:29.37 & 61:55:46.66 & LMM 49 & LkHA 200   & K3Ve& 13.47 & 0.29 & 00 & --- & 14.020 & 0.755 & 1.125 & 00 &  \\
04 & 00:42:34.46 & 61:54:01.13 & LMM 55 & ---        &     & 15.27 & 0.73 & 63 & --- & 15.522 &-0.007 & 0.640 & 27 &  \\
05 & 00:43:05.53 & 61:53:43.83 & LMM 77 & ---        & F8  & 13.52 & 0.65 & 90 & --- & 13.484 & 0.085 & 0.713 & 99 &  \\
06 & 00:43:07.27 & 61:46:27.41 & LMM 80 & ---        &     & 13.64 & 0.87 & 96 & --- & 13.600 &-0.003 & 0.877 & 09 &  \\
07 & 00:43:25.33 & 61:38:23.30 & LMM 91 & LkHA 201   & B2e & 14.32 & 1.29 & 00 & --- & 13.621 & 0.036 & 1.182 & 00 &  \\
08 & 00:43:25.65 & 61:48:51.68 & LMM 92 & ---        & A2  & 11.51 & 0.44 & 78 & B9IV& 11.560 & 0.120 & 0.420 & 43 &  \\
09 & 00:43:26.59 & 61:45:55.82 & LMM 94 & BD+60 87   & B9  & 10.25 & 0.62 & 93 & B8IV& 10.338 & 0.071 & 0.550 & 96 &  \\
10 & 00:43:28.87 & 61:48:04.10 & LMM 95 & ---        & A0  & 10.89 & 0.29 & 94 & A0IV& 10.940 & 0.135 & 0.250 & 42 &  \\
11 & 00:43:30.99 & 61:48:10.25 & LMM 96 & ---        & A7  & 12.22 & 0.58 & 69 & A5V & 12.172 & 0.090 & 0.551 & 92 &  \\
12 & 00:43:36.93 & 61:53:40.08 & LMM 98 & ---        & A5  & 12.04 & 0.36 & 94 & A4V & 11.982 & 0.156 & 0.347 & 67 &  \\
13 & 00:43:51.06 & 61:50:08.21 & LMM 113 & BD+61 157 & A0  & 10.64 & 0.24 & 94 & B9V & 10.630 &-0.020 & 0.280 & 49 &  \\
14 & 00:43:51.48 & 61:47:13.43 & LMM 114 & ---       & A1  & 10.92 & 0.20 & 86 & B9V & 10.890 & 0.040 & 0.230 & 69 &  \\
15 & 00:44:07.49 & 61:42:51.94 & LMM 123 & ---       &     & 13.63 & 0.93 & 94 & --- & 13.639 & 0.414 & 0.842 & 00 &  \\
16 & 00:44:12.80 & 61:51:01.89 & LMM 127 & ---       & A3  & 11.43 & 0.28 & 92 & A0V & 11.420 & 0.210 & 0.290 & 86 &  \\
17 & 00:44:16.55 & 61:50:44.05 & LMM 132 & ---       & A7  & 12.39 & 0.41 & 75 & A5V & 12.324 & 0.108 & 0.436 & 77 &  \\
18 & 00:44:20.76 & 61:49:45.15 & LMM 136 & ---       & F6  & 13.14 & 0.75 & 54 & --- & 13.116 & 0.120 & 0.720 & 01 &  \\
19 & 00:44:30.68 & 61:46:49.94 & LMM 149 & BD+60 94  & B8  &  9.73 & 0.17 & 94 & B8V & --- & --- & ---& --- & saturation \\
20 & 00:44:38.17 & 61:45:06.82 & LMM 157 & BD+60 95  & B9  & 10.23 & 0.19 & 00 & --- & 10.148 & -0.067& 0.183 & 03 &  \\
21 & 00:44:40.46 & 61:54:01.77 & LMM 161 & BD+61 162 & B9  &  9.67 & 0.14 & 93 & B9V & --- & --- & ---& --- & saturation \\
22 & 00:44:40.82 & 61:48:43.32 & LMM 163 & BD+61 163 & B6.5&  9.35 & 0.04 & 93 & B7IV-V& --- & --- & ---& --- & saturation\\
23 & 00:44:46.42 & 61:52:31.44 & LMM 170 & BD+61 164 & A0  & 10.03 & 0.08 & 75 & B9V  & 10.004 &-0.011 & 0.149 & 86 &  \\
24 & 00:44:47.45 & 61:56:49.16 & LMM 174 & ---       & A3  & 11.58 & 0.47 & 52 & A3IV & 11.982 & 0.156 & 0.347 & 67 &  \\
25 & 00:45:08.36 & 61:56:24.49 & LMM 197 & ---       & A7  & 12.52 & 0.55 & 71 & A7III-IV& 12.984&-0.246& 0.358 & 34 &  \\
26 & 00:45:18.73 & 61:46:17.77 & LMM 213 & ---       &     & 14.66 & 0.65 & 62 & --- & --- & --- & --- & --- & out of FoV \\
27 & 00:45:26.34 & 61:38:53.30 & LMM 219 & LkHA 205  &     & 14.39 & 1.49 & 00 & --- & --- & --- & --- & --- & out of FoV \\
28 & 00:45:49.70 & 61:42:43.63 & LMM 253 & ---       &     & 15.25 & 1.21 & 67 & --- & --- & --- & --- & --- & out of FoV \\
29 & 00:45:55.02 & 61:54:30.16 & LMM 260 & ---       & G5  & 14.61 & 1.60 & 77 & --- & --- & --- & --- & --- & out of FoV \\
30 & 00:46:03.23 & 61:52:17.76 & LMM 270 & ---       & G-K & 14.62 & 1.47 & 70 & --- & --- & --- & --- & --- & out of FoV \\
31 & 00:46:04.53 & 61:44:38.04 & LMM 271 & ---       &     & 14.32 & 0.76 & 51 & --- & --- & --- & --- & --- & out of FoV \\
\hline\hline
\end{tabular}
\end{center}
}
\end{table*}
\end{landscape}

The RVs of a considerable number  of stars in the direction of NGC~225
were measured for the first time in this study (Table 2). A mean RV of
$-8.3\pm 5.0$ km s$^{-1}$ was calculated  for the cluster. A search in
the literature  for stars  in Table  2 shows that  there are  only two
stars, whose RVs were previously measured. \cite{Meretal08} determined
the  RVs of  LMM~88 (BD+60~86)  and LMM~115  (BD+60~91) as  $-33.86\pm
0.76$~km~s$^{-1}$ and $-168.03\pm 0.26$~km~s$^{-1}$, respectively. Our
RV measurement for LMM~115 ($V_{r}=-168.2\pm 5.0$~km~s$^{-1}$) is in a
perfect agreement  with that  of \cite{Meretal08}, while  an agreement
between the two studies could not be  found for LMM~88 
($V_{r}=-129.0\pm 5.0$~km~s$^{-1}$) as this star is likely a 
spectroscopic binary.

In order to determine the galactic  orbit of the cluster,  we adopted
means of  the galactic orbital parameters found  for the cluster stars, 
whose  RVs were obtained currently from their spectra, as the galactic 
orbital parameters of NGC~225. 
The maximum  vertical distance  from the galactic  plane, eccentricity
and  period for  the  galactic  orbit of  NGC~225  were calculated  as
$Z_{max}= 90\pm 70$~pc, $e=0.07\pm  0.01$ and $P_{orb}= 255\pm 5$~Myr,
respectively.   \cite{Wuetal2009}   calculated  these   parameters  as
$Z_{max}= 70\pm  30$~pc, $e=0.09\pm 0.01$ and  $P_{orb}= 227.7\pm 4.5$
Myr  which are  in  agreement  with our  estimates.   We obtained  the
cluster's  apogalactic and  perigalactic  distances as  $R_{a}=9.37\pm
0.15$ and  $R_{p}=8.22\pm 0.12$~kpc,  respectively.  We  also computed
the space velocity components of the  cluster using the stars in
Table~6.  The space velocity components  of NGC~225 with respect to the 
galactic centre were determined as
$(U,V,W)=(16.02\pm 5.41, 219.75.25\pm 5.17, 0.82\pm 5.55)$~km~s$^{-1}$
in this study, where the  $V$  component for  galactic  rotation
velocity  of the  Sun ($V=220$  km s$^{-1}$) was used as the correction 
term.  Then, the  total space velocity of  the cluster  is estimated  
$S_{tot}=220$~km~s$^{-1}$. \cite{Wuetal2009}  calculated the uncorrected 
space velocity components of NGC~225 as $(U,V,W)=(38.0\pm 3.3, 209.6\pm
2.1, 6.6\pm  0.4)$ km s$^{-1}$  which give  a total space  velocity of
$S_{tot}=213$ km  s$^{-1}$.  These calculations show  that our results
about the  galactic orbit of the  cluster are in agreement  with those
found by \cite{Wuetal2009}.

\section{Acknowledgments}
Authors are grateful to the anonymous referees for their useful comments 
and improvements for the manuscript. This work has been supported in 
part by the Scientific and Technological Research Council (T\"UB\.ITAK) 
113F270. Part of this work was supported by the Research Fund of the 
University of Istanbul, Project Numbers: 39170 and 39582. We thank to 
T\"UB\.ITAK National Observatory for a partial support in using T100 
and RTT150 telescopes with project numbers 11BT100-180, 15AT100-738 
and 09BRTT150-470. We also thank to the on-duty observers and members 
of the technical staff at the T\"UB\.ITAK National Observatory for 
their support before and during the observations. This research has 
made use of the  SIMBAD and  ``Aladin sky atlas'' developed at CDS, 
Strasbourg Observatory, France, WEBDA and NASA\rq s Astrophysics Data 
System Bibliographic Services. We made use of data from the UVES 
Paranal Observatory project (ESO DDT Program ID 266.D-5655).


\begin{thebibliography}{}
\bibitem[Ak et al.(2016)]{Aketal2016}
Ak, T., Bostanc{\i}, Z.F., Yontan, T., Bilir, S., G\"uver, T., 
Ak, S.,  \"Urg\"up, H., \& Paunzen, E. 2016, Ap\&SS, 361, 126

\bibitem[Anders et al.(2004)]{Anders2004}
Anders, P., Bissantz, N., Fritze-v Alvensleben, U., \& de Grijs, R. 2004, 
MNRAS, 347, 196

\bibitem[Andreuzzi et al.(2002)]{Andetal2002}
Andreuzzi, G., Richer, H.B., Limongi, M., \& Bolte, M. 
2002, A\&A, 390, 961

\bibitem[Bagnulo et al.(2003)]{Bagnulo03}
Bagnulo S., Jehin E., Ledoux C., Cabanac R., Melo C., Gilmozzi R., \& ESO 
Paranal Science Operations Team 2003, The Messenger, 114, 10

\bibitem[Balaguer-N\'unez et al.(1998)]{Bala98}
Balaguer-N\'unez, L., Tian, K.P., \& Zhao, J.L. 1998, A\&AS, 133, 387

\bibitem[Becker \& Fenkart(1971)]{BeFen71}
Becker, W., \& Fenkart, R. 1971, A\&AS, 4, 241	

\bibitem[Bertin \& Arnouts(1996)]{BertArn1996}
Bertin, E., \& Arnouts, S. 1996, A\&AS, 117, 393

\bibitem[Bertin(2011)]{BertArn2011} 
Bertin, E. 2011, ASPC, 442, 435 

\bibitem[Bilir et al.(2010)]{Bilir10}
Bilir, S., G{\"u}ver, T., Khamitov, I., Ak, T., Ak, S., Co{\c s}kuno{\u g}lu, K.~B., 
Paunzen, E., \& Yaz, E. 2010, Ap\&SS, 326, 139  

\bibitem[Bilir et al.(2012)]{Bil12}
Bilir, S., Karaali, S., Ak, S., \" Onal, \" O., Da\u gtekin, N. D., Yontan, T., Gilmore, G., \& Seabroke, G.M. 2012, MNRAS, 421, 3362

\bibitem[Bostanc\i\ et al.(2015)]{Bosetal2015}
Bostanc{\i}, Z.F., Ak, T., Yontan, T., Bilir, S., G\"uver, T., Ak, S., 
\c{C}ak{\i}rl{\i}, \"O., \"Ozdarcan, O., Paunzen, E., De Cat, P., 
Fu, J.N., Zhang, Y., Hou, Y., Li, G., Wang, Y., Zhang, W., Shi, J., \& 
Wu, Y. 2015, MNRAS, 453, 1095

\bibitem[Bressan et al.(2012)]{Bresetal2012}
Bressan, A., Marigo, P., Girardi, L., Salasnich, B., Dal Cero, C.,
Rubele, S., \& Nanni, A. 2012, MNRAS, 427, 127

\bibitem[Brid\'zius et al.(2008)]{Brid2008}
Brid\u{z}ius, A., Narbutis, D., Stonkut\'e, R., Deveikis, V., \& 
Vansevi\u{c}ius, V. 2008, BaltA, 17, 337

\bibitem[Chen et al.(2014)]{Chenetal2014}
Chen, Y., Girardi, L., Bressan, A., Marigo, P., Barbieri, M., \& Kong, X. 
2014, MNRAS, 444, 2525

\bibitem[Co\c skuno\u glu et al.(2011)]{Coetal11}
Co\c skuno\u glu, B., et al. 2011, MNRAS, 412, 1237

\bibitem[Co\c skuno\u glu et al.(2012)]{Cosetal12}
Co\c skuno\u glu, B., Ak, S., Bilir, S., Karaali, S., \" Onal, \" O.,
Yaz, E., Gilmore, G., \& Seabroke, G. M. 2012, MNRAS, 419, 2844

\bibitem[Cox(2000)]{Cox2000}
Cox, A.N. 2000. Allen's astrophysical quantities, 4th ed.
Publisher: New York: AIP Press; Springer, 2000. Edited by
Arthur N. Cox. ISBN: 0387987460

\bibitem[de Meulenaer et al.(2013)]{deMeule2013}
de Meulenaer, P., Narbutis, D., Mineikis, T., \& Vancevi\u{c}ius, V. 
2013, A\&A, 550, 20

\bibitem[Dinescu et al.(1999)]{Dinetal1999}
Dinescu, D. I., Girardi, T.M., \& van Altena, W. F. 1999, AJ, 117, 1792

\bibitem[ESA(1997)]{ESA97}
ESA, 1997, The Hipparcos and Tycho catalogues. Publisher: Noordwijk, Netherlands: 
ESA Publications Division 1997, Series: ESA SP Series vol no: 1200, 
ISBN: 9290923997

\bibitem[Hoag et al.(1961)]{Hoagetal61}
Hoag, A.A., Johnson, H.L., Iriarte, B., Mitchell, R.I., Hallam, K.L., 
\& Sharpless, S. 1961, Publications of the U.S. Naval Observatory, 
2d ser., 17, 344

\bibitem[Hoag \& Applequist(1965)]{HoApp65}
Hoag, A.A., \& Applequist, N.L. 1965, ApJS, 12, 215

\bibitem[Janes \& Hoq(2013)]{Janes2013}
Janes, K., \& Hoq, S. 2013, AJ, 141, 92

\bibitem[Javakhishvili et al.(2006)]{Java06}
Javakhishvili, G., Kukhianidze, V., Todua, M., \& Inasaridze, R. 2006, A\&A, 447, 915

\bibitem[Johnson et al.(1961)]{Johetal61}
Johnson, H.L., Hoag, A.A., Iriarte, B., Mitchell, R.I., \& Hallam, K.L. 1961,  
Lowell Observatory, Bulletin No. 113, V.5, No.8, 133

\bibitem[Johnson \& Soderblom(1987)]{JS87}
Johnson, D.R.H., \& Soderblom, D.R. 1987, AJ, 93, 864

\bibitem[Karaali et al.(2003)]{Karaali03}
Karaali, S., Bilir, S., Karata{\c s}, Y., Ak, S.~G., 2003, PASA, 20, 165

\bibitem[Karaali et al.(2004)]{Karetal2004}
Karaali, S., Bilir, S., \& Hamzao\u glu, E. 2004, MNRAS, 355, 307

\bibitem[Karaali et al.(2005)]{Karaali05}
Karaali, S., Bilir, S., \& Tun{\c c}el, S. 2005, PASA, 22, 24

\bibitem[Karaali et al.(2011)]{Karetal2011}
Karaali, S., Bilir, S., Ak, S., Yaz, E., \& Co\c skuno\u glu, B. 2011, PASA, 28, 95

\bibitem[King(1962)]{King1962}
King, I. 1962, AJ, 67, 471

\bibitem[King et al.(2005)]{Kingetal2005}
King, I.R., Bedin, L.R., Piotto, G., Cassisi, S., \& Anderson, J. 
2005, AJ, 130, 626

\bibitem[Landolt(2009)]{Land2009}
Landolt, A.U. 2009, AJ, 137, 4186

\bibitem[Lang(2009)]{Lang09}
Lang, D. 2009, Thesis (Ph.D.), University of Toronto (Canada), ISBN: 9780494610008

\bibitem[Lattanzi et al.(1991)]{LMM91}
Lattanzi, M.G., Massone, G., \& Munari, U. 1991, AJ, 102, 177

\bibitem[Lee(1926)]{Lee26}
Lee, O.J. 1926, MNRAS, 86, 645	

\bibitem[Leggett(1992)]{Leg92}
Leggett, S.K. 1992, ApJS, 82, 351

\bibitem[Mermilliod et al. (2008)]{Meretal08}
Mermilliod, J.C., Mayor, M., \& Udry, S. 2008, A\&A, 485, 303

\bibitem[Mihalas \& Binney(1981)]{MB81}
Mihalas, D., \& Binney, J. 1981, In: Galactic Astronomy, second ed. Freeman, San
Fransisco.

\bibitem[Mowlavi et al.(2012)]{Mowlavietal2012}
Mowlavi, N., Eggenberger, P., Meynet, G., Ekstr{\"o}m, S., Georgy, C.,
Maeder, A., Charbonnel, C., \& Eyer, L. 2012, A\&A, 541, A41

\bibitem[Roeser et al.(2010)]{Roesetal2010}
Roeser, S., Demleitner, M., \& Schilbach, E. 2010, AJ, 139, 2440

\bibitem[Skrutskie et al.(2006)]{Sktr2006}
Skrutskie, M.F., et al. 2006, AJ, 131, 1163

\bibitem[Subramaniam et al.(2006)]{SMK06}
Subramaniam, A., Mathew, B., Kartha, S.S., et al. 2006, BASI, 34, 315

\bibitem[Sung et al.(2013)]{Sungetal2013}
Sung, H., Lim, B., Bessell, M.S., Kim, J.S., Hur, H., Chun, M., \& 
Park, B. 2013, JKAS, 46, 103

\bibitem[Svolopoulos(1962)]{Svo62}
Svolopoulos, S.N. 1962, ApJ, 136, 788	

\bibitem[Tang et al.(2014)]{Tangetal2014}
Tang, J., Bressan, A., Rosenfield, P., Slemer, A., Marigo, P., Girardi, L., \& 
Bianchi, L. 2014, MNRAS, 445, 4287

\bibitem[van Leeuwen(2007)]{vLe07}
van Leeuwen, F. 2007, A\&A, 474, 653

\bibitem[Wu et al.(2009)]{Wuetal2009}
Wu, Z.-Yu, Zhou, X., Ma, J., \& Du, C.-H. 2009, MNRAS, 399, 2146

\bibitem[Yontan et al.(2015)]{Yonetal2015}
Yontan, T., Bilir, S., Bostanc{\i}, Z.F., Ak, T., Karaali, S., G\"uver, T., 
Ak, S., Duran, \c S., \& Paunzen, E. 2015, Ap\&SS, 355, 267 

\end{thebibliography}
\end{document}